\documentclass[12pt]{article}

\usepackage{amsmath}
\usepackage{amssymb}
\usepackage{amsfonts}
\usepackage{slashed}
\usepackage{mathrsfs}
\usepackage{dsfont}

\newcommand{\CLsing}{{}^{1}C_L}
\newcommand{\CRsing}{{}^{1}C_R}
\newcommand{\CLtrip}{{}^{3}C_L}
\newcommand{\CRtrip}{{}^{3}C_R}
\newcommand{\CLRsing}{{}^{1}C_{L/R}}
\newcommand{\CLRtrip}{{}^{3}C_{L/R}}

\begin{document}

\begin{titlepage}

\begin{flushright}
arXiv:1907.02490
\end{flushright}
\vskip 2.5cm

\begin{center}
{\Large \bf The Lorentz- and CPT-Violating Standard Model Extension in Chiral
Perturbation Theory}
\end{center}

\vspace{1ex}

\begin{center}
{\large Brett Altschul,\footnote{{\tt baltschu@physics.sc.edu}}
and Matthias R. Schindler\footnote{{\tt mschindl@mailbox.sc.edu}}}

\vspace{5mm}
{\sl Department of Physics and Astronomy} \\
{\sl University of South Carolina} \\
{\sl Columbia, SC 29208} \\
\end{center}

\vspace{2.5ex}

\medskip

\centerline {\bf Abstract}

\bigskip

Lorentz and CPT violation in hadronic physics must be tied to symmetry violations
at the underlying quark and gluon level. Chiral perturbation theory provides
a method for translating novel operators that may appear in the Lagrange density for
color-charged parton fields into equivalent forms for effective theories at the meson
and baryon levels. We extend the application of this technique to the study of
Lorentz-violating and potentially CPT-violating operators from the minimal standard
model extension. For dimension-4 operators, there are nontrivial relations between the
coefficients of baryon-level operators related to underlying quark and gluon operators
with the same Lorentz structures. Moreover, in the mapping of the dimension-3 operators
from the quark and gluon level to the hadron level (considered here for the first time),
many of the hadronic observables contain no new low-energy coupling constants at all,
which makes it possible to make direct translations of bounds derived using experiments
on one kind of hadron into bounds in a completely different corner of the hadronic sector.
A notable consequence of this is bounds (at $10^{-15}$--$10^{-20}$ GeV levels) on
differences $a^{\mu}_{B}-a^{\mu}_{B'}$ of Lorentz and CPT violation coefficients for
$SU(3)_{f}$ octet baryons that differ in their structure by the replacement of a single
valance $d$ quark by a $s$ quark. Never before has there been any proposal
for how these kinds of differences could be constrained.

\bigskip

\end{titlepage}

\section{Introduction}
\label{sec-intro}

Recent developments in our understanding of fundamental symmetry principles have
led to a great deal of interest in testing how well the symmetries that seem
to underlie the fundamental physics we currently understand---the standard model of particle physics
and the general theory of relativity for gravitation---have really been verified
experimentally. Particular attention has been paid to Lorentz symmetry and CPT,
because these symmetries can be given up without needing to abandon the general
structure of field theory. Other exotic possibilities (such as violations of
the spin-statistics relation) are even less well behaved, and it may not even be possible
to formulate completely self-consistent test theories for such possibilities.

The experimental discovery of any kind of really exotic new fundamental phenomena would
obviously be of singular importance, on par with the development of renormalizable quantum
field theories, which provided a comprehensive framework for the study of interacting
elementary particles. If Lorentz or CPT violation is ever found experimentally, the new result can
immediately be analyzed in the context of effective field theory (EFT), since an
effective field theory framework capable of incorporating these symmetry violations into
the description of standard model quanta has already been developed~\cite{ref-kost1,ref-kost2}.
This EFT, known as the standard model extension (SME), describes Lorentz violation, and then is
automatically capable of describing CPT violation as well---because a quantum field theory (QFT)
with a well-defined $S$-matrix that is not invariant under CPT cannot be invariant under
Lorentz symmetry either~\cite{ref-greenberg}.

Thanks to its generality, the SME has become the standard formalism used for parameterizing
the results of experimental Lorentz and CPT tests. Most reasonable test theories previously proposed for
use in explorations of how these symmetries might be broken have turned out to be special cases of the
SME. As an EFT, the SME really contains a potentially infinite hierarchy of Lorentz-violating
operators that can be constructed out of standard model fields. However, in many cases, attention
is restricted to the minimal SME (mSME), which contains only gauge invariant and renormalizable
operators in its action. The mSME is expected to describe most low-energy Lorentz- and CPT-violating
phenomena, and experimental verifications of these symmetries can usually be expressed most usefully as
bounds on the coupling constants of the mSME (of which there is a large but finite number). The
Lagrange density for the mSME looks qualitatively similar to the Lagrange density for the standard
model itself. The key difference is that the mSME operators do not need to be Lorentz scalars; each
Lorentz-violating term will have one or more free Lorentz indices, which is contracted with a constant
background vector that describes a preferred directional structure in spacetime. These background constants
are the parameters that can be bounded experimentally, and the current state of the art for such bounds
may be found summarized in~\cite{ref-tables}. The best bounds on different strains of Lorentz-violating
operators come from many different areas of experimentation---including astronomy, atomic physics, and
collider physics.

There are still significant challenges for the interpretation of experimental results in terms of
SME parameters. One of the most important ones is a challenge that is also present in analyses in
a conventional standard model context. Although there are additional subtleties when Lorentz and CPT
are potentially broken, there is a common basic issue that the fundamental parameters of the theory
are the coefficients of operators that are formed from the elementary fields, which do not necessarily
represent the quanta that are physically accessible at low energies. In particular, it is not so
easy to take the results of measurements made on hadrons---particles with residual strong interactions
mediated largely by the exchange of mesons and meson-like resonance states---and relate those to the
fundamental description in terms of color-charged fields that are capable of exchanging gluons.
The purpose of chiral perturbation theory ($\chi$PT)~\cite{ref-weinberg,ref-gasser1,ref-gasser2}
(and see~\cite{ref-scherer} for a pedagogical introduction to the subject)
is to bridge this gap between the descriptions at the hadron level and the quark and gluon level.

Previous work has introduced a number of SME operators for quarks~\cite{ref-kamand1,ref-kamand2} and
gluons~\cite{ref-noordmans2,ref-noordmans1} and used $\chi$PT methods to translate them into equivalent formulations
for mesons and baryons. Ref.~\cite{ref-noordmans1} also considered certain radiative corrections and
meson-exchange potentials. However, there has not previously been a complete treatment of
all the mSME operators for strongly-interacting fields that are amenable to $\chi$PT methods simultaneously.  Such a
treatment is our goal in this paper. This is actually a slightly less onerous
undertaking than it might initially appear, since any Lorentz violation in nature is known to be a
very small effect. That means that it is a pretty much universally valid approximation to work only
to first order in the SME parameters; we shall not consider any operators or phenomena that involve
products of multiple SME coefficients. However, even at linear order, there are some interesting
relationships to be found between the coefficients.

The outline of this paper is as follows.
In section~\ref{sec-LV-QCD}, we introduce mSME Lorentz violation for the
fields at the level of two-flavor quantum chromodynamics (QCD)---the quarks and gluons. The methodology of
$\chi$PT is discussed in section~\ref{sec-chiral}. Then, in sections~\ref{sec-LV-meson}
and~\ref{sec-LV-baryon}, we construct the leading order (LO) effective actions
for the pion and nucleon sectors, respectively.
Experimental consequences, including some involving kaons and other strange particles, are discussed in
section~\ref{sec-expt}. Finally, section~\ref{sec-concl} summarizes our conclusions and areas
for future study.

\section{Lorentz Violation with QCD Fields}
\label{sec-LV-QCD}

\subsection{Quark Operators}
The starting point for our analysis will be the mSME Lagrange density, expressed in terms
of the QCD fields.
The mSME action is built out of gauge-invariant operators of dimensions 2, 3, and 4,
which are constructed out of the standard model's quantum fields. This is the same basic approach taken
in the usual standard model, except that the new operators specific to the mSME will have
free Lorentz indices. These indices are contracted with constant background
tensors; if the Lorentz violation arises through spontaneous symmetry breaking, then the background
tensors are determined from the vacuum expectation values of tensor-valued bosonic fields.
In the presence of such background tensors, otherwise identical experiments done in different
coordinate reference frames may yield different outcomes. By comparing the results of experiments
done with the apparatus at different orientations, or moving with different velocities, it is
possible to place bounds on the symmetry-breaking backgrounds.

The Lagrange density for the QCD sector of the mSME has operators that can be constructed
out of quark field bilinears and the gluon field.
Our focus will primarily be on Lorentz violation in two-flavor QCD.
However, when it is straightforward to do so, will we present generalizations to the theory containing a strange ($s$)
quark field, in addition to up ($u$) and down ($d$), with an approximate $SU(3)_{f}$ flavor symmetry. However, the inclusion
of a heavier quark does significantly increase the complexity of the theory, because there are no gauge symmetries
to prevent there existing a large number of Lorentz-violating mixing terms between the $d$ and $s$ fields.
The situation is analogous to having not just a single Cabbibo angle to describe the difference between the
mass eigenstates and electroweak eigenstates of the quarks, but a potentially different mixing angle for every
single component of the Lorentz-violating background tensors.

Moreover, although the focus of our analyses will always be the strongly-interacting sector
of the mSME, we will also make use of results from other sectors of the theory.
In addition to chiral symmetry and the $SU(3)_{c}$ gauge symmetry of QCD, there
are additional symmetry requirements that the hadronic Lagrangians must respect. Some of
these are simply the additional electroweak gauge symmetries of the standard model. However, there
are also other conditions that will need to be satisfied if the mSME (which is a QFT)
is to be embedded into a larger geometric
theory that also encompasses gravitation. We will employ these
additional consistency conditions freely, whenever they can be used to simplify the analysis.

We may further subdivide the various forms of Lorentz violation into those which are
odd under CPT, versus those that are CPT invariant. In the mSME, the
CPT-violating operators are those with odd numbers of Lorentz indices to be
contracted with the external background tensors. The CPT-even operators are
then those with even numbers of free indices; these include, naturally enough, the regular
standard model operators, which posses zero free Lorentz indices. (This rule---that whether
an operator is CPT violating can be determined simply by counting its indices---holds for most
operators in the full SME. However, there is an important exception~\cite{ref-colladay2,ref-altschul8}---the
$f$-type operators, which do not violate CPT, in spite of having odd numbers of indices.)
In the mSME, the only quark and gluon operators that can exist at mass dimension 4
are even under CPT. There are CPT-odd dimension-4 operators that can exist in a SME version
of pure quantum electrodynamics (QED), but all such operators involve Dirac matrix
structures that mix left- and right-chiral fermion fields in a way that is not consistent with the
$SU(2)_{L}$ electroweak gauge symmetry of the full standard model. Since these terms are not
gauge invariant (and are correspondingly not expected to be renormalizable), they are not truly
part of the mSME. However, similar terms that break the electroweak gauge symmetry actually
can exist as dimension-3
operators, where they may arise as vacuum expectation values of dimension-4 operators involving
the Higgs field. This is the same way that the Dirac fermion mass terms arise in the conventional standard
model; when the Higgs acquires a vacuum expectation, certain Yukawa-like dimension-4 operators
are converted into dimension-3 mass terms.

We shall first look at the dimension-4 operators, beginning with those for the quarks.
The CPT-even terms of this dimension that can exist in the quark sector are~\cite{ref-kost2}
\begin{equation}
\label{eq-qrk-cpt-even}
\mathcal{L}_{\rm quark}^{d\,=\,4,\,{\rm CPT-even}}=i(c_{Q})_{\mu\nu AB}\bar{Q}_{A}\gamma^{\mu}
D^{\nu}Q_{B}+i(c_{U})_{\mu\nu AB}\bar{U}_{A}\gamma^{\mu}D^{\nu}U_{B}
+ \,i(c_{D})_{\mu\nu AB}\bar{D}_{A}\gamma^{\mu}D^{\nu}D_{B}.
\end{equation}
The covariant derivatives contain all the standard model gauge fields, and in curved
spacetime, any derivatives must be taken as 50-50 linear combinations of derivative
operators acting to the right and left.
The left- and right-handed quark multiplets are denoted by
\begin{equation}
Q_{A}= 
\left[\begin{array}{c}
u_{A} \\
d_{A}
\end{array}
\right]_{L},
\quad U_{A} = \left[u_{A}\right]_{R},
\quad D_{A} = \left[d_{A}\right]_{R},
\end{equation}
where the left and right multiplets are of different dimensionalities because they transform
differently under the $SU(2)_{L}$ electroweak gauge symmetry.

The labels $A,B = 1,2,3$ denote the quark generations. Terms that are off diagonal in the $(A,B)$ basis
correspond to mixing between the generations due to Lorentz violation. It is familiar from the standard
model that there is generally not a single natural basis for the quark fields. The standard model is
formulated so that the mass terms in the quark Lagrangian are diagonal, so that there is no
flavor mixing during free quark propagation. However, the electroweak interactions are not diagonalized
in the quark mass basis, leading to flavor-changing interactions. In general, the Lorentz violation
coefficients will also not be diagonal in the mass basis. If all the heavier quarks are integrated out
of the theory via the renormalization group, leaving just the $u$ and $d$ fields, then the mixing issue
becomes moot. However, if the $s$ field is retained, then for each Lorentz component of the $(c_{Q})_{\mu\nu AB}$
and $(c_{D})_{\mu\nu AB}$, there are coefficients for unmixed $d$ and $s$ propagation, as well as a mixing
angle between them, analogous to the Cabbibo angle. As a result, the full parameter space of Lorentz-violating
flavor physics may be extremely difficult to probe, even with just three flavors.

The predominant effects of the dimension-4 Lorentz-violating operators are expected to come from terms
that are symmetric in the indices $(\mu,\nu)$. In particular, the antisymmetric parts
cannot modify the dimension-4 kinetic terms for baryons at leading order in the Lorentz violation, and
they cannot affect the dimension-4 kinetic operators for mesons at all. The
generic mSME Lagrange density for a single species of fermion is
\begin{eqnarray}
\label{eq-fermion-L}
\mathcal{L}_{{\rm spin-}\frac{1}{2}} & = & \bar{\psi}(i\mathit{\Gamma}^{\mu}\partial_{\mu}-M)\psi \\
\mathit{\Gamma}^{\mu} & = & \gamma^{\mu}+c^{\nu\mu}\gamma_{\nu}+d^{\nu\mu}\gamma_{5}\gamma_{\nu} \\
\label{eq-fermion-M}
M & = & m+im_{5}\gamma_{5}+a^{\mu}\gamma_{\mu}+b^{\mu}\gamma_{5}\gamma_{\mu}+\frac{1}{2}H^{\mu\nu}\sigma_{\mu\nu}.
\end{eqnarray}
With the only potential form of Lorentz violation coming from an
antisymmetric tensor $c^{\mu\nu}=-c^{\nu\mu}$, it is clear that the effect of $c^{\mu\nu}$ is, at leading order, just
a change in the basis of the Dirac matrices. A complementary transformation of the fermion field removes the antisymmetric
$c^{\mu\nu}$ from the Lagrange density at leading order~\cite{ref-colladay2}.
So the antisymmetric term cannot have any observable consequences
at leading order. The same fact can be seen manifested in the exact energy-moment relation for
a fermion described by $\mathcal{L}_{{\rm spin-}\frac{1}{2}}$ with just $c_{jk}\neq0$,
\begin{equation}
E=\sqrt{m^{2}+p_{j}p_{j}-2c_{jk}p_{j}p_{k}+c_{jl}c_{kl}p_{j}p_{k}}.
\end{equation}
In fact, it has been demonstrated that there is an exact supersymmetry transformation between 
$\mathcal{L}_{{\rm spin-}\frac{1}{2}}$ with just a $c^{\mu\nu}$ coefficient and the
general Lagrange density for a complex scalar field
\begin{equation}
\label{eq-spin0-general}
\mathcal{L}_{{\rm spin-0}}=\left(\partial^{\mu}-ia_{\phi}^{\mu}\right)\!\phi^{*}
\left(\partial_{\mu}+ia_{\phi}^{\mu}\right)\!\phi
+k^{\mu\nu}\partial_{\mu}\phi^{*}\partial_{\nu}\phi-m^{2}|\phi|^{2},
\end{equation}
so long as $a_{\phi}^{\mu}=0$ and
$k^{\mu\nu}=c^{\mu\nu}+c^{\nu\mu}+c^{\mu\rho}c^{\nu}\!{}_{\rho}$~\cite{ref-berger1}. [Note that it is not even
possible for the bosonic $k^{\mu\nu}$ to have an antisymmetric part without additionally breaking the
charge conjugation (C) symmetry of the Lagrange density.]

In the two-flavor QCD limit, the Lagrange density simplifies quite a bit.
Each of the $c_{\mu\nu}$ parameters in (\ref{eq-qrk-cpt-even}) is a dimensionless coupling constant,
and they form matrices which are Hermitian in the $(A,B)$ flavor space.
Restricting the Lagrange density of (\ref{eq-qrk-cpt-even}) to one with just $u$ and $d$ fields,
it reduces to
\begin{equation}
\label{eq-re-written-lag}
\mathcal{L}_{\rm light\, quarks}^{d\,=\,4,\,{\rm CPT-even}} 
=i\bar{Q}_{L} C_{L\mu\nu} \gamma^{\mu} D^{\nu} Q_{L}
+ i\bar{Q}_{R} C_{R\mu\nu}  \gamma^{\mu} D^{\nu} Q_{R},
\end{equation}
where the quark fields are now $Q_{L/R}=[u_{L/R},d_{L/R}]^T$, and the Lorentz-violation coefficients can be
collected in the matrices
\begin{equation}
C_{L/R}^{\mu\nu}=\left[
\begin{array}{cc}
c^{\mu\nu}_{u_{L/R}} & 0\\
0 & c^{\mu\nu}_{d_{L/R}}
\end{array}
\right].
\end{equation}
Note that there is no mixing between the $u$ and $d$ quarks; that is forbidden by
the standard model's unbroken electromagnetic gauge invariance.
This formalism actually allows for there to be different coefficients $c_{u_{L}}^{\mu\nu}$ and
$c_{d_{L}}^{\mu\nu}$, whereas in actuality, $SU(2)_{L}$ gauge invariance requires these to be
equal, $c_{u_{L}}^{\mu\nu}=c_{d_{L}}^{\mu\nu}=c_{q_{L}}^{\mu\nu}$. However, this is somewhat
modified when the $s$ quarks are included, and we shall generally consider the $c_{u_{L}}^{\mu\nu}$ and
$c_{d_{L}}^{\mu\nu}$ separately.

Because the coefficients in (\ref{eq-re-written-lag}) are given in the chiral basis, they multiply
operators that are not simply even or odd under parity (P) and C. Since most precision
experiments will measure effects that are unambiguously odd or even under P, the resulting bounds are
usually quoted on the linear combination $c^{\mu\nu}=\frac{1}{2}(c_{L}^{\mu\nu}+c_{R}^{\mu\nu})$
and $d^{\mu\nu}=\frac{1}{2}(c_{L}^{\mu\nu}-c_{R}^{\mu\nu})$. When dealing with hadrons and chiral symmetry,
it is often convenient to use different linear combinations of coefficients, broken up by their
transformation properties under isospin. The isosinglet is $\CLRsing^{\mu\nu}=
\frac{1}{2}\mathrm{Tr}(C_{L/R}^{\mu\nu})$,
and the isotriplet is $\CLRtrip^{\mu\nu}=C^{\mu\nu}_{L/R}-\mathds{1}\CLRsing^{\mu\nu}$,
where $\mathds{1}$ is the identity in isospin space.

There are also dimension-3 quark operators. Note that in the generic $\mathcal{L}_{{\rm spin-}\frac{1}{2}}$,
the di\-men\-sion-3 terms from (\ref{eq-fermion-M}) exhaust all the possible Dirac matrix
structures; each dimension-3 Lorentz-violating operator is composed of a fermion bilinear $\bar\psi{\cal A}\psi$,
multiplied by a matching background tensor. At dimension 4, some of the Dirac bilinear quantities
$\bar\psi{\cal B}\partial\psi$ were forbidden by electroweak gauge invariance. However, at dimension 3, terms
that mix left- and right-chiral fields can arise as vacuum expectations; in the standard model, this is
precisely how the mass $m$ appears. 
Among the allowed dimension-3 fermion bilinears in (\ref{eq-fermion-L}), there are two mass terms,
parameterized by $m$ and $m_{5}$.
We shall operate under the assumption that the $m_{5}$ has already been transformed away, so there
are only pure Dirac mass terms $m_{u}$ and $m_{d}$ in the two-flavor QCD Lagrange density. The way these
masses (which break chiral symmetry) are encoded in the hadronic sector will provide us with a guide for
how to include additional Lorentz-violating terms that may also softly break chiral invariance.

The softest breaking is by terms that are CPT odd,
\begin{equation}
\label{eq-qrk-cpt-odd}
\mathcal{L}_{\rm light\, quarks}^{d\,=\,3,\,{\rm CPT-odd}} 
=-\bar{Q}_{L} A_{L\mu} \gamma^{\mu} Q_{L}
-\bar{Q}_{R} A_{R\mu}  \gamma^{\mu}Q_{R},
\end{equation}
where the $A_{L/R}^{\mu}$ have a flavor-space matrix structure analogous to the $C_{L/R}^{\mu\nu}$:
\begin{equation}
A_{L/R}^{\mu}=\left[
\begin{array}{cc}
a^{\mu}_{u_{L/R}} & 0\\
0 & a^{\mu}_{d_{L/R}}
\end{array}
\right].
\end{equation}
Bounds on mSME coupling constants are usually expressed in terms of the vector $a^{\mu}$
and axial vector $b^{\mu}$ linear combinations,
\begin{equation}
\label{eq-qrk-a-b}
a^{\mu}_{u/d}=\frac{1}{2}(a_{{u_{L}/d}_{L}}^{\mu}+a_{u_{R}/d_{R}}^{\mu}), \quad b^{\mu}_{u/d} =
\frac{1}{2}(a_{{u_{L}/d}_L}^{\mu}-a_{{u_{R}/d}_R}^{\mu}).
\end{equation}
These also have isosinglet and
isotriplet linear combinations analogous to $\CLRsing^{\mu\nu}$ and $\CLRtrip^{\mu\nu}$.
In terms of these combinations, (\ref{eq-qrk-cpt-odd}) can be rewritten as
\begin{eqnarray}
\label{eq-qrk-cpt-odd-2}
\mathcal{L}_{\rm light\, quarks}^{d\,=\,3,\,{\rm CPT-odd}} & = & -\bar{Q}_{L} \left[ {}^{3\!\!}A_{L\mu} + \frac{1}{2}\left(
{}^{1\!\!}A_{R\mu} + {}^{1\!\!}A_{L\mu} \right)\mathds{1}\right]\gamma^{\mu} Q_{L} \\
 & &-\,\bar{Q}_{R} \left[{}^{3\!\!}A_{R\mu} + \frac{1}{2}\left( {}^{1\!\!}A_{R\mu} + {}^{1\!\!}A_{L\mu} \right)
\mathds{1}\right]\gamma^{\mu}Q_{R} \nonumber\\
& & - \,\bar{Q} \frac{1}{2}\left( {}^{1\!\!}A_{L\mu} - {}^{1\!\!}A_{R\mu} \right)\gamma_{5}\gamma^\mu  Q, \nonumber
\end{eqnarray}
which shows that this term includes an isosinglet axial vector current. This form of the Lagrange
density is particularly convenient when mapping to $\chi$PT.

Following the pattern of (\ref{eq-fermion-M}), there is one remaining possibility for $d=3$ operators---those
of the $H^{\mu\nu}$ type. Like the mass terms $m$ and $m_{5}$, the $H^{\mu\nu}$ Lorentz violation
mixes the left- and right-chiral fields directly, so the $H^{\mu\nu}$ do not need to have the kind of
natural chiral decomposition that the other SME terms possess. In fact, the
antisymmetry of $H^{\mu\nu}$ terms essentially preclude them making contributions to the LO $\chi$PT
Lagrange density, and so we will have little to say about these operators here.

\subsection{Gluon Operators}

There are also mSME operators in the purely gluonic sector. As in the quark sector, the dimension-4 gluon
operators are
even under CPT. In a strictly Minkowski spacetime, there is also an CPT-odd operator with mass dimension 3,
but this runs into difficulties when the EFT is embedded in a gravitometrodynamic theory such as
general relativity. This will ultimately mean that the CPT-even terms are the only ones that will need to
be considered.

Those CPT-even terms are collected in the form
\begin{equation}
\label{eq-LV-gluon-4}
{\cal L}^{d\,=\,4,\,{\rm CPT-even}}_{{\rm gluon}}=-\frac{1}{2}k_{G}^{\mu\nu\rho\sigma}
{\rm Tr}\left(G_{\mu\nu}G_{\rho\sigma}\right),
\end{equation}
with two powers of the gluon field strength tensor $G^{\mu\nu a}$. This allows for essentially arbitrary
bilinear products composed of spatial components of the chromoelectric and chromomagnetic fields, summed
symmetrically over the three colors. The four-index tensor $k_{G}^{\mu\nu\rho\sigma}$ has the symmetries of
the Riemann tensor and is double traceless. (A nonzero double trace would just provide a rescaling of
the usual QCD gluon Lagrange density.) Like the Riemann tensor, which can be broken into Ricci and Weyl parts,
the $k_{G}^{\mu\nu\rho\sigma}$ background can be split into two pieces with different characteristic behaviors,
\begin{equation}
k_{G}^{\mu\nu\rho\sigma}=\frac{1}{2}\left(\eta^{\mu\rho}\tilde{k}_{G}^{\nu\sigma}
-\eta^{\mu\sigma}\tilde{k}_{G}^{\nu\rho}
-\eta^{\nu\rho}\tilde{k}_{G}^{\mu\sigma}
+\eta^{\nu\sigma}\tilde{k}_{G}^{\mu\rho}\right)+\hat{k}_{G}^{\mu\nu\rho\sigma},
\end{equation}
where $\tilde{k}_{G}^{\mu\nu}=k_{G\alpha}\!{}^{\mu\alpha\nu}$ is symmetric, traceless in
$(\mu,\nu)$, and invariant under both C and PT. $\tilde{k}_{G}^{\mu\nu}$ is the gauge analogue of the
$c_{L/R}^{\mu\nu}$ terms for the chiral fermions. These terms represent there being different ``natural''
coordinates, which are oblique to the usual Cartesian coordinates, for the affected species.

While $\tilde{k}_{G}^{\mu\nu}$ is the ``Ricci-like'' part of the $k_{G}^{\mu\nu\rho\sigma}$ tensor, the
``Weyl-like'' part is $\hat{k}_{G}^{\mu\nu\rho\sigma}$. The two parts of the tensor have qualitatively
different features, and, in general $\hat{k}_{G}^{\mu\nu\rho\sigma}$ is expected to be less important in $\chi$PT.
There are two separate reasons for this. The first reason is that, because it has four free Lorentz indices,
any terms in the hadronic Lagrange density will need to involve either multi-particle interactions or
additional derivatives. In the mesonic sector, this immediately corresponds to terms that are higher order in
the chiral power counting. In the baryon sector, completely symmetrized combinations of the covariant baryon
derivatives can be included without a power counting penalty; however, the antisymmetry of the
$\hat{k}_{G}^{\mu\nu\rho\sigma}$ ensures that these terms vanish.

The second reason
is that the electromagnetic analogue of $\hat{k}_{G}^{\mu\nu\rho\sigma}$ is extremely tightly constrained.
The most important qualitative difference between the Ricci-like and Weyl-like tensors in the QED sector
of the mSME is that the ten Weyl-like terms generate photon birefringence, while the
nine Ricci-like components do not. The
birefringent terms can be bounded extremely well, by looking at photons that have traveled cosmological
distances---from radio galaxies, $\gamma$-ray bursts, and the cosmic microwave background. Some specific
linear combinations of these terms in the photon sector are constrained at the $10^{-37}$ level, and all the
birefringent terms are bounded at the $10^{-32}$ level, at least. This means that, in many contexts, it is
reasonable to neglect the birefringent electromagnetic terms entirely. The bounds on the Weyl-like gluonic
terms are not as strong as those for their electromagnetic equivalents. However, there will necessarily be
mixing between the different gauge sectors due to radiative corrections. A nonzero $\hat{k}_{G}^{\mu\nu\rho\sigma}$
will contribute to the renormalization of the birefringent photon terms; the mixing will be suppressed by
powers of the standard model coupling constants, but even with this modest suppression, the
$\hat{k}_{G}^{\mu\nu\rho\sigma}$ would need to be exceedingly small to be consistent with the
existing electromagnetic bounds.

The CPT-odd operator of dimension 3 has the form
\begin{equation}
\label{eq-LV-gluon-3}
{\cal L}^{d\,=\,3,\,{\rm CPT-odd}}_{{\rm gluon}}=k_{3}^{\mu}\epsilon_{\mu\nu\rho\sigma}
{\rm Tr}\left(G^{\sigma}G^{\nu\rho}+\frac{2}{3}ig_{s}G^{\sigma}G^{\nu}G^{\rho}\right).
\end{equation}
The electromagnetic analogue of this term will always generate birefringence, so it would also be
justifiable to neglect this term in any context in which $\hat{k}_{G}^{\mu\nu\rho\sigma}$
could be similarly neglected.

However, there is actually an even stronger reason to drop this term. The Lagrange
density in (\ref{eq-LV-gluon-3}) is not gauge invariant on its own. Instead, it changes by a total
derivative under a gauge transformation, provided the background tensor $k_{3}^{\mu}$ is a constant.
This means that the integrated action is gauge invariant, which is sufficient to ensure the equations
of motion are similarly gauge invariant. This is entirely satisfactory in a pure EFT approach in flat
spacetime. However, the physical mSME, if it is to represent the Lorentz and CPT violation that are
possible for real-world particles, must be embedded in a dynamical theory of gravitation. 
Explicit breaking of Lorentz invariance by constant vacuum tensors such as $k_{3}^{\mu}$ is inconsistent with a metric
theory of gravitation~\cite{ref-kost12}. Lorentz violation in a Riemannian theory of gravity is only possible
if the background tensors are themselves dynamical, with $k_{3}^{\mu}$ being determined by the vacuum expectation
value of a dynamical axial vector field; without this, the geometrical Bianchi identities cannot be satisfied.
Once there are nontrivial dynamics associated with $k_{3}^{\mu}$, ${\cal L}^{d\,=\,3,\,{\rm CPT-odd}}_{{\rm gluon}}$
no longer changes by a total derivative under a gauge transformation, meaning that the term is not
allowed, even in an asymptotically flat spacetime~\cite{ref-altschul37}. We shall not, therefore, consider
this term any further, although if it were included in the $\chi$PT Lagrangian, it would be coupled to hadrons
in the same way as a quark $b^{\mu}$ term.

\section{Elements of $\chi$PT}
\label{sec-chiral}

With the full quark and gluon Lagrange density set down, we now find ourselves in a position to construct a
new, effective Lagrange density for the hadrons.
Our analysis of how the Lorentz-violating mSME operators are to be embedded in $\chi$PT will begin with a
treatment of the purely mesonic Lagrangian. (Some qualitative results for pions can even be extended to
their octet partners with nonzero strangeness, especially to $K$ mesons.)
There can be a basically self-contained description
of the pions in $\chi$PT, without needing to simultaneously introduce nucleons. In contrast, a low-energy
$\chi$PT treatment of baryons automatically includes, in addition to a description of the
free propagation of the baryons, a set of meson-baryon interaction vertices.

Whichever baryon sector is under consideration, using $\chi$PT means
considering all possible terms that are permitted by the symmetries of the underlying
theory~\cite{ref-weinberg,ref-gasser1,ref-gasser2}.
Normally, in Lorentz-invariant QCD, this suite of symmetries includes
rotations, boosts, and the discrete transformations of C, P, and time reversal (T). There is
also an accidental chiral symmetry to QCD. This symmetry is exact when the quarks are massless, $m_{u}=m_{d}=0$,
and even when the masses are nonvanishing, the chiral transformations generate an approximate symmetry that
has many useful consequences at energy scales well below the symmetry breaking scale of $\sim4\pi F\approx 1$ GeV,
where $F\approx 92.4$ MeV is the pion decay constant.
The strongly interacting QCD dynamics break the full chiral symmetry group $SU(2)_L \times SU(2)_R$ down
to its diagonal subgroup $SU(2)_V$.\footnote{While it is common to refer to $SU(2)_L\times SU(2)_R$ as
the chiral symmetry group, the QCD Lagrange density for massless $u$ and $d$ quarks has a $U(2)_L\times U(2)_R$
symmetry. Because of the axial anomaly, this reduces to a $SU(2)_L\times SU(2)_R \times U(1)_V$ symmetry,
with the $U(1)_V$ symmetry related to baryon number. This will be relevant for the baryonic sector.}
The pions are the associated pseudo-Goldstone bosons;
in the $m_{u}=m_{d}=0$ limit, in which the original chiral symmetry is exact, the pions are
precisely massless.

The massless, two-flavor QCD Lagrange density
\begin{equation}
\label{eq-masslessQCD}
\mathscr{L}^0_{\rm QCD} = \bar{Q}_L i \slashed{D} Q_L + \bar{Q}_R i \slashed{D} Q_R -
\frac{1}{2}{\rm Tr}(G_{\mu\nu}G^{\mu\nu}).
\end{equation}
will be the starting point for $\chi$PT. (We are continuing to follow our previous
convention~\cite{ref-kamand1,ref-kamand2} of using the letter
variants $\mathscr{L}$ for Lorentz-invariant Lagrange densities and ${\cal L}$ for Lorentz-violating ones.)
In (\ref{eq-masslessQCD}),
$Q_{L/R}=[u_{L/R},d_{L/R}]^T$ are the doublets of left- and right-chiral quark fields; and
$D_\mu q = (\partial_\mu + i g G_\mu) q$ is the QCD covariant derivative, with $G_\mu$ the
gluon fields, $g$ the strong coupling constant, and $G_{\mu\nu}$ the gluon field strength tensor.
If (\ref{eq-masslessQCD}) is the entire Lagrange density (that is, if the $u$ and $d$ masses, along with any
other sources of explicit chiral symmetry breaking, are vanishing), then there are global symmetry
transformations,
\begin{equation}
Q_L \rightarrow L Q_L,\quad Q_R \rightarrow R Q_R,
\end{equation}
where $(L,R)$ are a pair of matrices in $SU(2)_{L} \times SU(2)_{R}$.

However, since this chiral symmetry is broken down to $SU(2)_V$, there are Goldstone modes.
The Goldstone boson fields carry the quantum numbers of the broken symmetry
generators. This means that pion fields can be encoded in the $SU(2)$ matrix~\cite{ref-coleman1}
\begin{equation}
\label{eq-U}
U(x) = \exp\left[ i \frac{\phi(x)}{F} \right].
\end{equation}
Here, $\phi = \sum \phi_a\tau_a$ [so that the $\phi$ contains the three $SU(2)$ generators],
and $F$ is the pion decay constant in the $SU(2)$ chiral limit.
Global chiral transformations act on $U(x)$ as
\begin{equation}
\label{eq-U-transform}
U(x)\rightarrow U'(x)= RU(x)L^{\dagger},
\end{equation}
for $(L,R)\in SU(2)_{L} \times SU(2)_{R}$.

The effective action for the pure pion EFT (the lowest-energy limit of QCD) can be constructed from
the matrix $U(x)$ and its derivatives. The power counting scheme used in $\chi$PT dictates that
each additional derivative acting on a pion field indicates an additional power of a small parameter; this
applies to both spatial and temporal derivatives, because the pion mass is small in the chiral limit. The
lowest-order chirally-invariant term that can be constructed out of $U(x)$ contains the meson kinetic terms.
The standard LO pion Lagrange density thus has a term of the form
\begin{equation}
\label{eq-lo-chpt}
\mathscr{L}_{\pi}^{\rm LO}\supset \frac{F^2}{4} {\rm Tr}(\partial_\mu U \partial^{\mu} U^\dagger),
\end{equation}
where the trace Tr is taken over flavor space.

However, in real-world QCD, the masses of the light quarks cannot usually be so glibly neglected.
Moreover, in addition to gluon interactions, there are also interactions between the quarks and
the electroweak gauge fields. Both of these facts can be included in the $\chi$PT in a unified way,
by treating the quark masses and the electroweak gauge boson fields as external fields. These
external fields are included in the QCD Lagrange density in the form
\begin{equation}
\label{eq-QCD-external}
\mathscr{L} =  \mathscr{L}^0_{\rm QCD} + \mathscr{L}_{\rm external},
\end{equation}
in which the coupling to the external fields is described by~\cite{ref-gasser1,ref-gasser2}
\begin{eqnarray}
\label{eq-LagExtLC}
\mathscr{L}_{\rm external} & =  & \bar{Q}_L \gamma^\mu \left(l_\mu+\frac{1}{3}v_\mu^{(s)}\mathds{1}\right)Q_L
+ \bar{Q}_R \gamma^\mu \left(r_\mu+\frac{1}{3}v_\mu^{(s)}\mathds{1} \right)Q_R \\
& & +\, \bar{Q}_L (s-ip)Q_R + \bar{Q}_R (s+ip)Q_L. \nonumber 
\end{eqnarray}
The external fields $l_{\mu}$, $r_{\mu}$, $s$, and $p$ can have nontrivial
structures in flavor space. As chiral fields, $l_{\mu}$ and $r_{\mu}$ may be taken to be traceless [the trace part
of the Lagrange density being taken care of through the isosinglet term $v_{\mu}^{(s)}$; no axial vector
singlet term is needed because the diagonal chiral symmetry is broken at a higher energy scale by the
chiral anomaly] and thus
represented in terms of the generators
\begin{equation}
l^{\mu} = \frac{1}{2}\sum\tau_{a}l^{\mu}_{a},\quad r^{\mu} = \frac{1}{2}\sum\tau_{a}r^{\mu}_{a} .
\end{equation}
With appropriate choices, these can give the couplings of the quarks to the electroweak gauge bosons. For instance,
setting just $l_{3}^{\mu}=r_{3}^{\mu}=v^{(s)\mu}=-\frac{1}{2}eA^{\mu}$ to be nonzero gives the vector couplings of the $u$ and $d$
quarks to the electromagnetic four-vector potential $A^{\mu}$.
[The combinations including $v_{\mu}^{(s)}$ as they appear in (\ref{eq-LagExtLC}), which are also frequently
useful, can be denoted $\tilde{l}_{\mu}=l_{\mu}+\frac{1}{3}v_{\mu}^{(s)}\mathds{1}$ and
$\tilde{r}_{\mu}=r_{\mu}+\frac{1}{3}v_{\mu}^{(s)}\mathds{1}$.]

The Dirac mass terms for the $u$ and $d$ fields can be
introduced similarly, through the scalar external field
$s={\cal M}={\rm diag}\,(m_u,m_d)$. [The pseudoscalar $p$ could be used for Majorana masses like
$m_{5}$ in~(\ref{eq-fermion-M}).] All of the external fields break the chiral symmetry, so
the form that this symmetry breaking takes must be mirrored between the Lagrange densities at the
QCD level and hadron level. To match the symmetry breaking patterns it is necessary to determine
how the external fields would need to transform if (\ref{eq-QCD-external}) were actually to remain
chirally invariant.
In fact, the Lagrange density (\ref{eq-QCD-external}) is invariant under not just a global chiral
transformation, but a local $(V_L,V_R,{\cal U})\in SU(2)_L\times SU(2)_R\times U(1)_{V}$,
\begin{equation}
\label{eq-local-transform}
Q_L \rightarrow \exp\left[-\frac{i\Theta(x)}{3}\right]V_L(x) Q_L
, \quad Q_R \rightarrow \exp\left[-\frac{i\Theta(x)}{3}\right]V_R(x) Q_R,
\end{equation}
so long as the external fields transform as
\begin{eqnarray}
l_\mu & \rightarrow & V_L l_\mu V_L^\dagger + i V_L \partial_\mu V_L^\dagger \nonumber\\
r_\mu & \rightarrow & V_R r_\mu V_R^\dagger + i V_R \partial_\mu V_R^\dagger \nonumber \\
\label{eq-ext-transform}
v_\mu^{(s)} & \rightarrow & v_\mu^{(s)}-\partial_{\mu}\Theta\\
s+ip & \rightarrow & V_R (s+ip) V_L^\dagger \nonumber \\
s-ip & \rightarrow & V_L (s-ip) V_R^\dagger. \nonumber
\end{eqnarray}
The $\Theta(x)$ is associated with the $U(1)_{V}$ baryon number symmetry, which is separate from the
chiral $SU(2)_L\times SU(2)_R$.
The invariance under \emph{local} chiral transformations ensures that the chiral Ward identities are
satisfied~\cite{ref-gasser1,Leutwyler:1993iq}.
With the quark mass terms transforming as $s$, (\ref{eq-ext-transform}) implies the transformation
behavior $\mathcal{M} \rightarrow V_R \mathcal{M} V_L^\dagger$.

At the hadronic level, the particle excitations may also have nonminimal couplings to external fields.
To get the minimal couplings, we ensure invariance under local
chiral transformations by replacing the derivative $\partial_{\mu}U$ of $U(x)$ by a covariant derivative
with a chiral connection,
\begin{equation}
\label{eq-pion-cov}
D_{\mu}U=\partial_{\mu}U+iUl_{\mu}-ir_{\mu}U.
\end{equation}
This transforms under local transformations according to
$D_\mu U \rightarrow V_R D_\mu U V_L^\dagger$. Then the possible nonminimal couplings can be constructed from
the ``field strengths'' formed out of the chiral connection fields $l^{\mu}$ and $r^{\mu}$,
\begin{eqnarray}
f_L^{\mu\nu} & = & \partial^\mu l^\nu - \partial^\nu l^\mu - i [l^\mu, l^\nu] \\
f_R^{\mu\nu} & = & \partial^\mu r^\nu - \partial^\nu r^\mu - i [r^\mu, r^\nu].
\end{eqnarray}
These transform covariantly under the local transformations,
\begin{equation}
f_L^{\mu\nu} \rightarrow V_L f_L^{\mu\nu} V_L^\dagger, \quad
f_R^{\mu\nu} \rightarrow V_R  f_R^{\mu\nu} V_R^\dagger.
\end{equation}

The mass enters in a similar fashion, via the external field
\begin{equation}
\chi=2B(s+ip),
\end{equation}
transforming as $\chi\rightarrow V_R \chi V_L^\dagger$.
The constant $B$ is numerically determined by the nontrivial dynamics of strong-field QCD. However, it can
be directly related to the chiral condensate density, $B=-\frac{1}{2}\langle\bar{Q}Q\rangle$. Thus
the full LO pion Lagrangian, including nonzero quark masses and the couplings to external fields, is given
by~\cite{ref-gasser2}
\begin{equation}
\label{eq-LC-pion-LO}
\mathscr{L}_\pi^{\rm LO} = \frac{F^2}{4}{\rm Tr}(D_\mu U D^\mu U^\dagger) +
\frac{F^2}{4}{\rm Tr}(\chi U^\dagger + U \chi^\dagger).
\end{equation}
This provides a relationship, $M_\pi^2 =-\frac{1}{2}\langle\bar{Q}Q\rangle(m_u + m_d)/F^{2}$,
between the pion mass and the underlying quark masses.
(Although the quark masses are real, $\chi^{\dag}$ is still formally distinguished from $\chi$ in this situation.)

For the various quantities that can be used to assemble the mesonic Lagrange densities, the power counting
scheme is
\begin{equation}
U= {\cal O}(q^0), \quad D^\mu U = {\cal O}(q), \quad \chi = {\cal O}(q^2), \quad f_{L/R}^{\mu\nu} = {\cal O}(q^2),
\end{equation}
where $q$ is a small momentum expansion parameter.

For the baryonic sector, which resides at a slightly higher natural momentum scale than the minimal meson theory,
there are additional quantities that can be invoked in the construction of chirally invariant Lagrange densities.
The starting point is the nucleon doublet $\Psi = [p,n]^T$, which transforms
as~\cite{ref-coleman1,ref-callan1,ref-georgi}
\begin{equation}
\Psi \rightarrow K(V_L,V_R,U) \Psi,
\end{equation}
with the matrix $K(V_L,V_R,U)$ determined in terms of the transformation rules for the square root $u(x)$ of $U(x)$.
If $[u(x)]^{2}=U(x)$, then in order to have $u(x)\rightarrow\sqrt{V_{R}UV^{\dag}_{L}}$, the matrix $u(x)$ itself
must transform according to
\begin{equation}
\label{eq-KDef}
u(x) \rightarrow V_R u K^\dagger= Ku V_L^\dagger.
\end{equation}

For the baryon field $\Psi$, the chiral covariant derivative is more complicated than the one
(\ref{eq-pion-cov}) for the pions. Probably most notably, the covariant derivative that acts on
the fermions includes not just the external fields, but also the meson fields themselves, which enter
through combinations of $u(x)$ and $u^{\dag}(x)$,
\begin{equation}
\label{eq-chiral-conn}
\Gamma_{\mu}=\frac{1}{2}\left[u^\dagger(\partial_\mu-ir_\mu)u+u(\partial_\mu - il_\mu)u^\dagger\right],
\end{equation}
so that
\begin{equation}
\label{eq-NucCovDer}
D_\mu \Psi =\left[\partial_\mu + \Gamma_\mu - i v_\mu^{(s)}\mathds{1} \right]\Psi.
\end{equation}
This covariant derivative is constructed so that $D_{\mu}\Psi$ transforms in the same way as $\Psi$ itself,
$D_{\mu}\Psi\rightarrow KD_{\mu}\Psi$.

In addition to a kinetic coupling term involving $D_{\mu}\Psi$, it is well known that the nucleon also has
an axial vector coupling term. With this term included, the Lorentz-invariant LO pion-nucleon Lagrangian
has the form~\cite{Gasser:1987rb}
\begin{equation}
\label{eq-piNLag}
\mathscr{L}_{\pi N}^{\rm LO} = \bar{\Psi}\left( i\slashed{D} - m +\frac{g_A}{2}
\gamma^\mu\gamma_5 u_\mu\right) \Psi.
\end{equation}
In this equation, $m$ is the nucleon mass and $g_A$ the axial coupling, both in the chiral limit. At LO,
these may be replaced by their physical values of $m_N \approx 939\text{ MeV}$ and $\mathtt{g}_A \approx 1.27$,
although there are further corrections to the physical values at higher chiral orders.
The chiral vielbein $u_{\mu}$ is defined as
\begin{equation}
\label{eq-vielbein}
u_{\mu}=i\left[ u^\dagger (\partial_\mu-ir_\mu) u - u (\partial_\mu - il_\mu) u^\dagger \right],
\end{equation}
which transforms according to $u_\mu \rightarrow Ku_\mu K^\dagger$.

Because the nucleon mass $m_{N}$ does not vanish in the chiral limit, a timelike derivative
acting on the nucleon field will not be suppressed, even at low energies. This affects the
chiral $q$-counting scheme. The additional building blocks
defined in the nucleon sector are counted as
\begin{equation}
\Psi = {\cal O}(q^0), \quad D_\mu \Psi = {\cal O}(q^0), \quad u_\mu = {\cal O}(q).
\end{equation}
However, because $\Psi$ must obey a field equation, the particular
combination $(i\slashed{D} - m_N)\Psi$ is counted as ${\cal O}(q)$. This means, for instance,
that $\slashed{D}\Psi$ may be exchanged for $-im_{N}\Psi$ if terms of higher chiral orders
are being neglected~\cite{ref-fettes}.

\section{Lorentz-Violating Mesonic Lagrange Density}
\label{sec-LV-meson}

\subsection{CPT-Even Operators}

The Lagrange density (\ref{eq-LC-pion-LO})
can be generalized in a straightforward way to include Lorentz violation coming
from the quark and gluon sectors. We shall begin with generalizations to
the kinetic Lagrange density (\ref{eq-lo-chpt}). The results with just the dimension-4
quark terms have already been given~\cite{ref-kamand1}.
The argument that led to these terms was based on matching the transformation properties of the QCD-level
Lagrange density (\ref{eq-re-written-lag}) onto the equivalent meson-scale Lagrange density.
Under a chiral transformation with matrices $(L,R)$, the doublets of $u$ and $d$ quark fields transform as
$Q_{R}\to RQ_{R}$ and $Q_{L}\to LQ_{L}$. This takes (\ref{eq-re-written-lag})
\begin{equation}
{\cal L}_{\rm light quarks}^{d\,=\,4,\,{\rm CPT-even}} \to
i\bar{Q}_{L} L^{\dagger} C_{L\mu\nu} L \gamma^{\mu} D^{\nu} Q_{L}
+ i\bar{Q}_{R} R^{\dagger} C_{R\mu\nu} R \gamma^{\mu} D^{\nu} Q_{R}.
\end{equation}
With constant matrices $C_{L/R}^{\mu\nu}$ that do not transform under $SU(2)_{L} \times SU(2)_{R}$,
the presence of the Lorentz-violating term (\ref{eq-re-written-lag}) would break the chiral symmetry.
However, if the $C_{L/R}^{\mu\nu}$ were also to transform,
\begin{equation}
\label{eq-CLR-trans}
C_L^{\mu\nu}   \to L C_L^{\mu\nu} L^{\dagger},\quad 
C_R^{\mu\nu}  \to R C_R^{\mu\nu} R^{\dagger},
\end{equation}
the chiral symmetry would be restored. Since the transformation properties (\ref{eq-CLR-trans}) would keep
the quark-level Lagrange density chirally invariant, applying those same transformation
prescriptions must also maintain the chiral symmetry at the hadron level. This rule allows us to
identify what kinds of operators the $C_{L/R}^{\mu\nu}$ can be associated with in the pion Lagrange density.

The transformation properties (\ref{eq-CLR-trans}) are more usefully expressed in terms of the isospin
singlet and triplet components of the $C_{L/R}^{\mu\nu}$. The isosinglet is useful because
it does not transform at all under chiral rotations, while the isotriplet retains
the transformation properties of the underlying $C_{L/R}^{\mu\nu}$.

Moreover, along with the $C_{L/R}^{\mu\nu}$,
which modify the kinetic terms in the quark Lagrange density, there is also the gluon $\tilde{k}_{G}^{\mu\nu}$,
which---since it appears in a term (\ref{eq-LV-gluon-4}) involving only the gauge fields---also does not
transform at all under the action of the chiral $SU(2)_{L} \times SU(2)_{R}$. So the transformation
rules for the coefficients of the dimension-4 operators are
\begin{eqnarray}
\label{eq-iso-CLR-trans}
\CLsing\rightarrow\CLsing, & & \CLtrip\rightarrow L\CLtrip L^{\dagger}, \\
\CRsing\rightarrow\CRsing, & & \CRtrip\rightarrow R\CRtrip R^{\dagger}, \nonumber\\
\tilde{k}_{G}^{\mu\nu} & \rightarrow & \tilde{k}_{G}^{\mu\nu}. \nonumber
\end{eqnarray}

These transformation rules---(\ref{eq-CLR-trans}) or (\ref{eq-iso-CLR-trans}), along with the discrete
transformation properties of the SME terms---are sufficient for us to
determine the qualitative forms of the operators these coefficients are associated with in the
LO mesonic Lagrangian. The process begins with writing down all the possible operator forms
that are consistent with the chiral symmetry. However,
the Lorentz-violating terms in the quark-level Lagrange density are also the only potential sources of C,
P, and T violations in the theory. So at LO, any terms in the hadronic Lagrange densities need to
have the same discrete symmetries as the terms in the underlying quark density that are multiplied
by the same SME coefficients. This means that the coefficients for left- and right-handed
quark fields must always enter the pion Lagrange density multiplied by the same low-energy couplings
(LECs). In this way, imposing the discrete symmetries drastically reduces the number of independent terms
in the Lagrangian. Moreover, a number of the remaining terms turn out to be linearly
dependent (or at least linearly dependent at LO). Using integration by parts, the additional redundant
terms may also be eliminated from the description of the theory.

The LO minimal mesonic Lagrange density is given by
\begin{eqnarray}
\label{eq-lo-pionlagrangian}
\mathcal{L}_{\pi}^{d\,=\,4,\,{\rm LO}} & = & \left[\beta^{(1)}\left(^{1}C_{R\mu\nu} +\,^{1}C_{L\mu\nu}\right)+\beta^{(3)}
\tilde{k}_{G\mu\nu}\right]\frac{F^2}{4}
{\rm Tr}\left[(D^{\mu}U)^{\dagger}D^{\nu}U\right] \\
& & +\beta^{(2)}\frac{F^2}{4} \mathrm{Tr}\left[(D^{\mu}U)^{\dagger}\, ^{3}C_{R\mu\nu}D^{\nu}U +
D^{\mu}U\, ^{3}C_{L\mu\nu} (D^{\nu}U)^{\dagger}\right] \nonumber
\end{eqnarray}
where the $\beta^{(n)}$ are dimensionless LECs. (The ``$d=4$'' superscript denotes the mass dimension of the
operators in the underlying QCD Lagrange density that give rise to this mesonic expression, rather than
the dimension of the ${\cal L}_{\pi}^{d\,=\,4,\,{\rm LO}}$ operators themselves.)
The factor of $F^2/4$ in (\ref{eq-lo-pionlagrangian})
is present to mirror the form of the standard pion Lagrange density and is also chosen such
that based on naive dimensional analysis~\cite{ref-manohar}, the 
$\beta^{(n)}$ are expected to have a natural size that is ${\cal O}(1)$. Actually, the $\beta^{(2)}$ term
does not contribute at all at leading order. It was shown in~\cite{ref-kamand1} that with symmetric tensors
$\CLRtrip^{\mu\nu}$, the $\beta^{(2)}$ reduces to a total derivative. As we shall see below, this
actually holds for antisymmetric $\CLRtrip^{\mu\nu}$ as well.

The short-distance QCD physics is entirely encapsulated in the LECs. A complete determination of their values
would entail the use of nonperturbative QCD, and to our knowledge,
no numerical computation of these values has thus far been
undertaken. Relative to the formulation given in~\cite{ref-kamand1,ref-kamand2}, the portion of
(\ref{eq-lo-pionlagrangian}) that is symmetric in $(\mu,\nu)$ contains one additional term, since in addition
to the four quark tensors $c^{\mu\nu}_{u_{L}}$, $c^{\mu\nu}_{d_{L}}$, $c^{\mu\nu}_{u_{R}}$, and $c^{\mu\nu}_{d_{L}}$,
(\ref{eq-lo-pionlagrangian}) also includes the contribution from the gluon tensor
$\tilde{k}_{G}^{\mu\nu}$~\cite{ref-noordmans1}. However, it turns out that, when all five of these tensors
from the mSME are included, there is actually a nontrivial relation between the LECs, which will allow us to
express $\beta^{(3)}$ in terms of $\beta^{(1)}$.

What the $c^{\mu\nu}$ and $\tilde{k}_{G}^{\mu\nu}$ tensors represent is a form of Lorentz violation in which
the natural spacetime coordinates for different standard model fields are actually different. Having solely a
nonzero $c^{\mu\nu}_{u_{L}}$, for example, indicates that the left-chiral $u$ quarks propagate according to
normal relativistic rules in a coordinate system that is oblique to the usual coordinates. If we change to the oblique
coordinates, which are given (at leading order) by $x'^{\mu}=x^{\mu}-\frac{1}{2}(c_{u_{L}})^{\mu}\!{}_{\nu}x^{\nu}$, 
the dynamics for the $u$ quark field are standard, but all the other fields will have Lorentz-violating
behavior, dictated by $c'^{\mu\nu}=\frac{1}{2}\tilde{k}'^{\mu\nu}_{G}=-c^{\mu\nu}_{u_{L}}$ for the remaining species.
The fact that $c$-type Lorentz violation can be moved from one sector to another by coordinate redefinitions like
these means that any physical measurement of a $c$-type coefficient really has to be a measurement of a
difference of the coefficients for different particle types.

Expanding $U(x)$ to second order in the pion fields, the Lagrange density (\ref{eq-lo-pionlagrangian})
gives the Lorentz-violating kinetic terms for the pions. [Expanding $U(x)$ to higher orders in the pion
fields produces Lorentz-violating meson interaction vertices.]
The two-pion portion of the Lagrange density is
\begin{eqnarray}
\label{eq-lo-2pionlagrangian}
\mathcal{L}_{\pi}^{{\rm LO},\,2\phi} & = & \left[\frac{\beta^{(1)}}{4}(c^{\mu\nu}_{u_{L}}+c^{\mu\nu}_{d_{L}}+
c^{\mu\nu}_{u_{R}}+c^{\mu\nu}_{d_{L}})+\frac{\beta^{(3)}}{2}\tilde{k}_{G}^{\mu\nu}\right]
\partial_{\mu}\phi_{a}\partial_{\nu}\phi_{a} \\
& = & \left[\frac{\beta^{(1)}}{4}(c^{\mu\nu}_{u_{L}}+c^{\mu\nu}_{d_{L}}+
c^{\mu\nu}_{u_{R}}+c^{\mu\nu}_{d_{L}})+\frac{\beta^{(3)}}{2}\tilde{k}_{G}^{\mu\nu}\right] \\
& & \times(\partial_{\mu}\pi^{+}\partial_{\nu}\pi^{-} +
\partial_{\mu}\pi^{-}\partial_{\nu}\pi^{+} + \partial_{\mu}\pi^{0}\partial_{\nu}\pi^{0}). \nonumber
\end{eqnarray}
[$\pi^{0}=\phi_{3}$, $\pi^{+}=\frac{1}{\sqrt{2}}(\phi_{1}-\textit{i}\phi_{2})$, and $\pi^{-}=
\frac{1}{\sqrt{2}}(\phi_{1}+\textit{i}\phi_{2})$ are the physical pion fields.]
This $\mathcal{L}_{\pi}^{{\rm LO},\,2\phi}$
has the form of the Lorentz-violating $k$ term from (\ref{eq-spin0-general}). There are three
species of pions, but in the chiral limit, there is a just single
\begin{equation}
\label{eq-kpi}
k_{\pi}^{\mu\nu}=\frac{\beta^{(1)}}{2}
(c_{u_{L}}^{\mu\nu}+c_{u_{R}}^{\mu\nu}+c_{d_{L}}^{\mu\nu}+c_{d_{R}}^{\mu\nu})+\beta^{(3)}\tilde{k}_{G}^{\mu\nu}
\end{equation}
tensor common to all three of the physical pion fields. Note that since the pion wave functions
are all equal mixtures of left- and right-chiral, $u$ and $d$ quarks, the quark portion of
$k_{\pi}^{\mu\nu}$ receives equal contributions from each of the four quark types.

The nontrivial relation between $\beta^{(1)}$ and $\beta^{(3)}$ arises from the fact that, by making
a change of coordinates in the usual two-flavor QCD Lagrange density
$x^{\mu}\rightarrow x'^{\mu}=x^{\mu}+\kappa^{\mu}\!{}_{\nu}x^{\nu}$ (for some arbitrary symmetric tensor
$\kappa^{\mu\nu}$), we can turn the conventional QCD expression into a Lorentz-violating Lagrange
density with $c^{\mu\nu}_{u_{L}}=c^{\mu\nu}_{d_{L}}=c^{\mu\nu}_{u_{R}}=c^{\mu\nu}_{d_{L}}=
\frac{1}{2}\tilde{k}_{G}^{\mu\nu}=\kappa^{\mu\nu}$. Since the theory this describes is really just
the standard, Lorentz-invariant one, merely viewed in unconventional coordinates, the pion sector must
also be the usual one, expressed in the same oblique coordinates. This means that
$\frac{1}{2}k_{\pi}^{\mu\nu}=\kappa^{\mu\nu}$ also.

Taken together with (\ref{eq-lo-2pionlagrangian}), this relation indicates that $\beta^{(3)}=1-\beta^{(1)}$.
The same kind of relation for the $c$-type Lorentz-violation coefficients for composite particles was
found in~\cite{ref-altschul38}, with the coefficient for a composite being a sum of the constituents'
coefficients, each one weighted by the fraction of the total momentum carried by a particular
constituent. In this case, $\beta^{(1)}$ represents the fraction of the pion momentum
carried by all the constituent quarks, with the remainder carried by the gluons. The values of these
weights still cannot be determined without recourse to nonperturbative QCD, but (\ref{eq-kpi}) does
simplify to
\begin{equation}
k_{\pi}^{\mu\nu}=\frac{\beta^{(1)}}{2}
(c_{u_{L}}^{\mu\nu}+c_{u_{R}}^{\mu\nu}+c_{d_{L}}^{\mu\nu}+c_{d_{R}}^{\mu\nu})+\left[1-\beta^{(1)}\right]
\tilde{k}_{G}^{\mu\nu}.
\end{equation}
This specific results also supports the general presumption that each of the LECs should be ${\cal O}(1)$.

The $k_{\pi}^{\mu\nu}$ coefficients are the easiest ones to observe directly for pions. They affect
the energy-momentum relations for ultrarelativistic pions, which in turn can lead to new thresholds
(including upper energy thresholds) for reactions involving extremely energetic mesons. There are
also pion vertices, which are in some cases straightforward Lorentz-violating generalizations of
the usual pion vertex operators, involving even numbers of fields. The form of
(\ref{eq-lo-2pionlagrangian}) involves the insertion of a Lorentz-violating symmetric tensor between
the $(\mu,\nu)$ indices of the derivatives $\partial_{\mu}\phi_{a}\partial_{\nu}\phi_{a}$. At higher
orders in the fields $\phi_{a}$, there are homologous expressions, such as 
\begin{equation}
\label{eq-lo4pi}
\mathcal{L}_{\pi}^{{\rm LO},\,4\phi}=
\frac{k_{\pi}^{\mu\nu}}{6F^{2}}(\phi_{a}\phi_{b}\partial_{\mu}\phi_{a}\partial_{\nu}\phi_{b}
-\phi_{b}\phi_{b}\partial_{\mu}\phi_{a}\partial_{\nu}\phi_{a}) 
\end{equation}
at fourth order. Note that all these higher-order terms depend on the same linear combination of
quark and gluon SME coefficients.

Naively it looks like there might be other terms, associated with the antisymmetric parts of
$\CLRsing^{\mu\nu}$ and $\CLRtrip^{\mu\nu}$ or with the $H^{\mu\nu}$, which would
be qualitatively different in structure. (Note that, by virtue of its structure, $\tilde{k}_{G}^{\mu\nu}$
cannot have an antisymmetric part, so that the antisymmetric terms can only involve quark parameters.)
For example, if the $C_{L/R}^{\mu\nu}$ are all antisymmetric, then direct expansion of the Lagrange
density gives
\begin{equation}
\label{eq-asym-zero}
\mathcal{L}_{\pi}^{{\rm LO}}\supset
\frac{\beta^{(2)}}{4}\left(c_{u_{L}}^{\mu\nu}+c_{u_{R}}^{\mu\nu}+c_{d_{L}}^{\mu\nu}+c_{d_{R}}^{\mu\nu}\right)
(\partial_{\mu}\phi_{1}\partial_{\nu}\phi_{2}-\partial_{\nu}\phi_{1}\partial_{\mu}\phi_{2}).
\end{equation}
However, (\ref{eq-asym-zero}) is actually a total derivative (both with respect to $\partial_{\mu}$
and $\partial_{\nu}$), which makes no contribution to the physics.

We might also anticipate a three-$\phi$ term involving
$\partial_{\mu}\phi_{3}\partial_{\nu}\phi_{a}\phi_{a}-\partial_{\nu}\phi_{3}\partial_{\mu}\phi_{a}\phi_{a}$,
or equivalently,
$\partial_{\mu}\pi^{0}(\pi^{-}\partial_{\nu}\pi^{+}+\pi^{+}\partial_{\nu}\pi^{-})
-\partial_{\nu}\pi^{0}(\pi^{-}\partial_{\mu}\pi^{+}+\pi^{+}\partial_{\mu}\pi^{-})$.
However, not only would this term be another total derivative, but
the three-pion form gives an operator that is manifestly odd under C, which does not
match the symmetry of the SME coefficients multiplying the term; this C-odd behavior is a general
feature of antisymmetric tensor SME coefficients in scalar field theories~\cite{ref-altschul30}.
In fact, there appears to be no term that can be written
down in the pion sector at LO that involves an antisymmetric tensor structure. This observation was already
prefigured by the fact that there was no antisymmetric tensor among the external fields (\ref{eq-ext-transform}) that
could be coupled to the hadrons at leading order. This also justifies the absence of any LO terms involving
$\hat{k}_{G}$, which is separately antisymmetric in two sets of Lorentz indices.

\subsection{CPT-Odd Operators}
\label{sec-pion-CPTodd}

For the $d=3$, CPT-odd operators coming from the quark sector, finding their couplings to pions is
actually quite straightforward. These terms can simply be inserted as external fields of the left-
and right-chiral vector forms, through $-\tilde{l}^{\mu}$ and $-\tilde{r}^{\mu}$. The correct
signs and magnitudes
for these terms can be read off directly from the SME coupling (\ref{eq-qrk-cpt-odd})
[or equivalently (\ref{eq-qrk-cpt-odd-2})] to the
quarks. The pion term is then
\begin{eqnarray}
\label{eq-d3-pion}
{\cal L}_{\pi}^{d\,=\,3,\,{\rm LO}} & = & \frac{F^{2}}{4}
{\rm Tr}\left[(\partial^{\mu}U+iUA_{L}^{\mu}-iA_{R}^{\mu}U)
(\partial^{\mu}U+iUA_{L}^{\mu}-iA_{R}^{\mu}U)^{\dag}\right] \nonumber\\
& &-\,\frac{F^{2}}{4}{\rm Tr}(\partial_{\mu}U\partial^{\mu}U^{\dag}).
\end{eqnarray}
The scalar part with $v^{(s)}_{\mu}$ cancels between the left- and right-chiral terms,
which ensures that the expression  has the correct behavior under C and P transformations.
Moreover, (\ref{eq-d3-pion}) is structured to contain only Lorentz-violating terms, since
the usual LO meson kinetic term has been explicitly subtracted away.
In Lorentz-invariant $\chi$PT, the singlet axial-vector current is not considered, and
even in the SME, it is not possible to construct an axial vector current operator entirely out of
pseudoscalar meson fields.

Simplifying (\ref{eq-d3-pion}), and noting that $A_{L}^{\mu}+A_{R}^{\mu}=(a_{u}^{\mu}+a_{d}^{\mu})\mathds{1}+
(a_{u}^{\mu}-a_{d}^{\mu})\tau_{3}$, 
the CPT-odd expression reduces to
\begin{eqnarray}
{\cal L}_{\pi}^{d\,=\,3,\,{\rm LO}} & \supset & \frac{i}{4}{\rm Tr}\left[(A_{L}^{\mu}+A_{R}^{\mu})
(\phi_{a}\partial_{\mu}\phi_{b}-\partial_{\mu}\phi_{a}\phi_{b})\tau_{a}\tau_{b}\right] \\
\label{eq-pion-a}
& = & -\frac{i}{2}\left(a_{u_{L}}^{\mu}+a_{u_{R}}^{\mu}-a_{d_{L}}^{\mu}-a_{d_{R}}^{\mu}\right)
\left(\pi^{+}\partial_{\mu}\pi^{-}-\pi^{-}\partial_{\mu}\pi^{+}\right),
\end{eqnarray}
up to a total derivative.
The form of (\ref{eq-pion-a}) is essentially what is expected for a charged spin-0 field. Note that this kind of
term cannot exist for a single real scalar field, so the CPT-odd term does not
affect the $\pi^{0}$ part of the Lagrange density.
As far back as~\cite{ref-kost27}, it was argued that the net $a_{\phi}^{\mu}$ term for a meson should
be a difference of the $a$-type coefficients for the constituent quark fields, times a dimensionless
factor not too different from unity. This calculation grounds that conclusion firmly in $\chi$PT.
In fact, since the $a$-type terms are odd under C, but independent of spin and momentum, it makes
sense that the expectation value of the contribution from virtual quark-antiquark pairs to the net
meson $a_{\phi}^{\mu}$ should vanish.

Like the $c$-type coefficients, the $a^{\mu}$ coefficients
for fermions can only be observed as differences between different species, not in isolation. Moreover, the
difference must be between the coefficients for species that can interconvert. For example, in a theory
with multiple species of massless, noninteracting fermions, none of the $a_{L/R}^{\mu}$ can be observed by
propagation effects. The free propagation of a particle with SME coefficient $a^{\mu}$ and
momentum $p^{\mu}$ is indistinguishable from the motion of a particle with $a'^{\mu}=0$ and momentum
$p'^{\mu}=p^{\mu}-a^{\mu}$, and without the ability to create or annihilate particles, it is impossible to
make an absolute measurement of the momentum carried by an excitation.
Introducing a Dirac mass term generates a coupling between the left-
and right-chiral fermion modes, which makes differences of $a_{L}^{\mu}$ and $a_{R}^{\mu}$ physically
observable; these are precisely the fermion $b^{\mu}$ terms, which affect the energy-momentum relations
of massive particles in a directly observable fashion.

The reason that only differences between $a^{\mu}$ values are observable is tied to the observation that
$a^{\mu}$ effectively represents a translation of the momentum space for a single species. That translation
can be undone by applying a field redefinition~\cite{ref-colladay2} that changes the phase of the
fermion field by $e^{-ia\cdot x}$. For uncoupled species, the phases of their fields may be varied
independently. However, if two types of fermions are coupled by an interaction term of the form
$\bar{\psi}_{a}{\cal C}\psi_{b}$, then the phases of $\psi_{a}$ and $\psi_{b}$ cannot be set
separately; trying to define away both a $a_{a}^{\mu}$ and $a_{b}^{\mu}$ will leave behind a residual
term in the Lagrange density, proportional to $a_{a}^{\mu}-a_{b}^{\mu}$.

The combination $a_{\pi^{+}}^{\mu}=a_{u}^{\mu}-a_{d}^{\mu}=\frac{1}{2}(a_{u_{L}}^{\mu}+a_{u_{R}}^{\mu}-a_{d_{L}}^{\mu}
-a_{d_{R}}^{\mu})$ that appears in (\ref{eq-pion-a}) is thus not actually
yet an observable, since it is a difference of $a$-type parameters for two species ($u$ and $d$ quarks)
which do not have the same charge and thus cannot interconvert. In fact, to form a physical observable, we
must construct a difference of two $a$-type parameters for like-charged meson species. (There are possible
exceptions to this rule if the $a$-type coefficients are to be measured in a gravitational experiment; however,
even there, nonminimal gravitational couplings are required, placing this scenario outside the mSME
framework.) We shall return to
this topic in section~\ref{sec-expt}, when we discuss experimental bounds on CPT violation for mesons.

\section{Lorentz-Violating Baryonic Lagrange Density}
\label{sec-LV-baryon}

\subsection{CPT-Even Operators}

The analysis of the contributions from dimension-4 mSME operators in the nucleon sector proceeds along
similar lines to the treatment in the pion sector. Again, there is a straightforward generalization
of earlier results~\cite{ref-kamand1,ref-kamand2} to include the additional contributions from a
gluon $\tilde{k}_{G}^{\mu\nu}$ term. Because of the presence of chirally covariant derivatives, the
form of the free baryon Lagrange density also determines the LO meson-baryon couplings.

The LO baryonic Lagrange density for the nucleon doublet field $\Psi$ is
\begin{eqnarray}
\label{eq-lo-pionucleonlagr}
\mathcal{L}^{d\,=\,4,\,{\rm LO}}_{\pi N} & = &
\alpha^{(1)}\bar{\Psi}[(u^{\dagger}\;{^3C_{R}^{\mu\nu}} u + u\: {^3C_{L}^{\mu\nu}}
u^{\dagger})(\gamma_{\nu} i D_{\mu} +
\gamma_{\mu} i D_{\nu})]\Psi \\
& & +\,\alpha^{(2)}\left({^1C_{R}^{\mu\nu}} + \: {^1C_{L}^{\mu\nu}}\right)\bar{\Psi}(\gamma_{\nu}
i D_{\mu} +
\gamma_{\mu}i D_{\nu})]\Psi \nonumber\\
& & +\,\alpha^{(3)}\bar{\Psi}[(u^{\dagger}\;{^3C_{R}^{\mu\nu}} u - u\: {^3C_{L}^{\mu\nu}}
u^{\dagger})(\gamma_{\nu}\gamma^{5}
i D_{\mu} + \gamma_{\mu}\gamma^{5} i D_{\nu})]\Psi \nonumber\\
& & +\,\alpha^{(4)}\left({^1C_{R}^{\mu\nu}} - \:
{^1C_{L}^{\mu\nu}}\right)\bar{\Psi}(\gamma_{\nu}\gamma^{5}
i D_{\mu} + \gamma_{\mu}\gamma^{5} i D_{\nu})\Psi \nonumber\\
& & +\,\alpha^{(5)}\tilde{k}_{G}^{\mu\nu}\bar{\Psi}(\gamma_{\nu}iD_{\mu}+\gamma_{\mu}i D_{\nu})\Psi,
\nonumber
\end{eqnarray}
where the $\alpha^{(n)}$'s are the dimensionless LECs for this sector of the theory. By naive dimensional
analysis, these are again anticipated to be ${\cal O}(1)$. The structural properties of these various
terms are discussed in detail in~\cite{ref-kamand1}.

As there was
for the pions, there is a nontrivial constraint coming from the fact that, when all the
quark $c^{\mu\nu}_{u_{L}}=c^{\mu\nu}_{d_{L}}=c^{\mu\nu}_{u_{R}}=c^{\mu\nu}_{d_{L}}$ and gluon
$\frac{1}{2}\tilde{k}_{G}^{\mu\nu}$ are equal to $\kappa^{\mu\nu}$, the theory is really just
conventional QCD written in skewed coordinates. From the expression for the proton coefficient
\begin{equation}
\label{eq-cp}
c_{p}^{\mu\nu}=\alpha^{(1)}(c_{u_{L}}^{\mu\nu}+c_{u_{R}}^{\mu\nu}-
c_{d_{L}}^{\mu\nu}-c_{d_{R}}^{\mu\nu})
+\alpha^{(2)}(c_{u_{L}}^{\mu\nu}+c_{u_{R}}^{\mu\nu}+
c_{d_{L}}^{\mu\nu}+c_{d_{R}}^{\mu\nu})+2\alpha^{(5)}\tilde{k}_{G}^{\mu\nu},
\end{equation}
it again follows, from $c_{p}^{\mu\nu}=\kappa^{\mu\nu}$, that $\alpha^{(5)}=\frac{1}{4}-\alpha^{(2)}$.
[Precisely the same result could be obtained from the neutron coefficient $c_{n}^{\mu\nu}$,
because the $\alpha^{(1)}$ term, which changes sign between protons and neutrons, vanishes
when all the quark coefficients are equal.] So in
spite of the inclusion of the additional gluonic SME coefficients relative to~\cite{ref-kamand1},
the number of independent LECs corresponding to the $d=4$ QCD operators has not increased.

\subsection{CPT-Odd Operators}
\label{sec-baryon-cpt-odd}

The LO contributions from the CPT-violating vector and axial vector operators enter through their
couplings to the chiral connection~(\ref{eq-chiral-conn}). Here, in order to get the correct C and
P transformation properties, we must set the chiral sources
$l_{\mu}= -{}^{3\!\!}A_{L\mu}$ and
$r_{\mu}= -{}^{3\!\!}A_{R\mu}$.
In addition, from comparing (\ref{eq-qrk-cpt-odd-2}) and (\ref{eq-LagExtLC}), we see that
\begin{equation}
v_{\mu}^{(s)}=-\frac{3}{2}\left({}^{1\!\!}A_{L}^{\mu}+{}^{1\!\!}A_{R}^{\mu}\right)
=-\frac{3}{4}
\left(a_{u_{L}}^{\mu}+a_{u_{R}}^{\mu}+a_{d_{L}}^{\mu}+a_{d_{R}}^{\mu}\right).
\end{equation}
Inserting these into the chiral covariant derivative gives
\begin{equation}
\label{eq-bary-cov-deriv}
(D_{\mu}-\partial_{\mu})\Psi=
\frac{1}{2}\left\{u^\dagger(\partial_{\mu}+i\,{}^{3\!\!}A_{R\mu})u
+u(\partial_{\mu}+i\,{}^{3\!\!}A_{L\mu})u^\dagger
+ i\frac{3}{2}\left[{}^{1\!\!}A_{L\mu}+{}^{1\!\!}A_{R\mu}\right]\mathds{1}\right\}\Psi.
\end{equation}
There is also the axial coupling term, which likewise depends on $l_{\mu}$ and $r_{\mu}$,
\begin{equation}
\label{eq-bary-vielbein}
\frac{g_{A}}{2}\gamma^{\mu}\gamma_{5}u_{\mu}=i\frac{g_{A}}{2}\gamma^{\mu}\gamma_{5}
\left[u^\dagger\left(\partial_{\mu}+i\,{}^{3\!\!}A_{R\mu}\right)u
-u\left(\partial_{\mu}+i\,{}^{3\!\!}A_{L\mu}\right)u^\dagger\right].
\end{equation}
In addition, we need to include the singlet axial vector contribution from the quark-level
Lagrange density. While chiral symmetry does not constrain this piece of the interaction
and thus provides no relationships between various terms with different numbers of pion fields,
only the contribution without pions will be relevant for the following discussion. The corresponding
baryonic operator takes the form
\begin{equation}
\mathcal{L}^{d\,=\,3}_{N}\supset -\alpha^{(6)}\bar{\Psi}\gamma_{5}\gamma^{\mu}\left({}^{1\!\!}A_{L\mu} -
{}^{1\!\!}A_{R\mu}\right)\mathds{1}\Psi,
\end{equation}
where $\alpha^{(6)}$ is a new LEC. (If we had considered hadronic terms arising from
the CPT-odd gluon operator with coefficient $k_{3}^{\mu}$, they would also have entered here, through yet
another $\bar{\Psi}\gamma_{5}\gamma^{\mu}\Psi$ operator with another new LEC.)

So, with the neglect of the pion coupling terms [setting $u(x)=1$] the CPT-violating part of the
purely baryonic action reads
\begin{eqnarray}
\mathcal{L}^{d\,=\,3}_{N} & = & \bar{\Psi}\Big\{\gamma_{\mu}\frac{1}{2}\left[
-\left({}^{3\!\!}A_{L}^{\mu}+{}^{3\!\!}A_{R}^{\mu}\right)
-3 \left({}^{1\!\!}A_{L}^{\mu}+{}^{1\!\!}A_{R}^{\mu}\right)\mathds{1}\right]
-\frac{g_{A}}{2}\gamma_{5}\gamma_{\mu}\left({}^{3\!\!}A_{L}^{\mu}-{}^{3\!\!}A_{R}^{\mu}\right)  \nonumber\\
& & -\,\alpha^{(6)}\gamma_{5}\gamma_{\mu}\left( {}^{1\!\!}A_{L}^{\mu} - {}^{1\!\!}A_{R}^{\mu} \right) \mathds{1}
\Big\}\Psi.
\end{eqnarray}
From this, coefficients such as the proton $a^\mu$ and $b^\mu$ can be read off,
\begin{eqnarray}
\label{eq-proton-a}
a_{p}^{\mu} & = & \left(a^{\mu}_{u_{L}}+a^{\mu}_{u_{R}}\right)
+\frac{1}{2}\left(a_{d_{L}}^{\mu}+a_{d_{R}}^{\mu}\right)
=2a_{u}^{\mu}+a_{d}^{\mu}\\
\label{eq-proton-b}
b_{p}^{\mu} & = & \frac{g_{A}}{4}\left(a^{\mu}_{u_{L}}-a^{\mu}_{u_{R}}\right)+\frac{\alpha^{(6)}}{2}
\left(a^{\mu}_{u_{L}}-a^{\mu}_{u_{R}}+a^{\mu}_{d_{L}}-a^{\mu}_{d_{R}}\right)
=\frac{g_{A}}{2}b_{u}^{\mu}+\alpha^{(6)}\left(b_{u}^{\mu}+b_{d}^{\mu}\right).
\end{eqnarray}
Since $b_{p}^{\mu}$ is directly observable, it is a sum of direct differences between
the $a$-type coefficients for pairs of equally charged chiral species. Moreover, while $a_{p}^{\mu}$
is not an independent physical observable, it has a very natural form---the sum of the
(spin-averaged) $a$-type coefficients for the proton's three valance quarks. It is actually
quite remarkable that, at LO, there is only a single undetermined LEC (which only affects the
baryons' $b$-type coefficients, not any of the $a$-type coefficients) that appears in the dimension-3
Lagrange densities for both the pions and the nucleons.

\section{Experimental Constraints}
\label{sec-expt}

We shall now turn to an exploration of how the various LECs for mesons and baryons can be constrained
using existing and future experimental data. In purely phenomenalistic analyses, it has been
commonplace to assign a separate set of SME coefficients to each observable hadron species. However,
this will end up significantly over-counting the number of independent parameters, because the true
number of mSME coefficients for strongly interacting particles
is determined by the structure of the quark and gluon sectors. The coefficients for different types
of hadrons are not independent, and this makes it possible to carry bounds over from one part of the
strongly interacting sector to another. There will be modest uncertainties, due to the presence of
unknown LECs; however, it will be possible to set constraints on the SME parameters for baryons using
measurements made on mesons, and vice versa. This is one of the things that makes $\chi$PT such a
powerful technique.

We have previously discussed~\cite{ref-kamand1} how bounds on pion Lorentz violation could be improved
by making reference to atomic clock experiments that measured Lorentz violation for nucleons,
and~\cite{ref-noordmans1} took a similar approach to constraining the gluon coefficients
$\tilde{k}_{G}^{\mu\nu}$. $\chi$PT methods can also be used to help isolate Lorentz-violating observables
in the weak sector~\cite{ref-kamand2}. All these approaches have dealt with the dimension-4, CPT-even
coefficients. Since this paper has, for the first time, given a $\chi$PT description of dimension-3,
CPT-odd operators for quarks, gluons, and hadrons, we shall primarily concentrate our attention on how
new bounds may be placed on these dimension-3 operators.

However, we should first point out that the specific bounds derived in~\cite{ref-kamand1} were set
under the simplifying assumption that there was no dimension-4 Lorentz
violation in the gluon Lagrange density. In that case, particular sums of proton and neutron observables
ended up probing the exact same linear combinations
$c_{u_{L}}^{\mu\nu}+c_{u_{R}}^{\mu\nu}+c_{d_{L}}^{\mu\nu}+c_{d_{R}}^{\mu\nu}$ as
a separate set of pion observables (in the chiral limit). Meanwhile,~\cite{ref-noordmans1} adopted a
complementary approach, effectively assuming that there was Lorentz violation in the gluon
sector, and none for the quarks. If, as discussed here, all the phenomenalistically viable dimension-4
QCD operators are included, the actual effective coefficients for mesons and baryons are linear
combinations of elements from the quark and gluon Lagrange densities, and the relative weights
for the two kinds of coefficients are not known. As a result, bounds such as those
derived in~\cite{ref-kamand1,ref-noordmans1} should be considered order of magnitude estimates for
the sizes of the underlying quark and gluon SME coefficients; the bounds (at the $10^{-19}$--$10^{-27}$
levels) represent the largest those
coefficients could be without there being unnatural fine tuning in the form of a
nearly exact cancelation between the quark and gluon parameters. 

We now turn to the experimental status of the dimension-3 hadronic terms.
In many cases, the $b$-type coefficients for nucleons are extremely well bounded. The reason is that
the $b^{\mu}$ coefficients alter the energies of spin states, meaning that these coefficients
can be measured in extremely sensitive spin flip and spin precession experiments. Except for the
proton time component $b_{p}^{T}$, all the components of $b_{p}^{\mu}$ and $b_{n}^{\mu}$ have been
bounded at the $10^{-25}$ GeV level or better~\cite{ref-tables}. Bounds on forms of Lorentz violation
are by convention expressed in a system of Sun-centered celestial equatorial coordinates $(T,X,Y,Z)$,
with the $Z$-axis coinciding with the Earth's rotation axis. The $X$- and $Y$-components of
a vector such as $b_{p}^{\mu}$ are relatively easy to constrain, because they affect observables that
oscillate as the Earth rotates; bounds on a $Z$-component are trickier, since while such a component does
give rise to anisotropic phenomena, they are not of a type that can be observed just by looking for sidereal
variations in some observable; and measuring a time component is the hardest, as it requires a direct
test of either boost invariance or a discrete symmetry. This explains why $b_{p}^{T}$ has, thus far, only
been bounded at the $3\times10^{-8}$ level~\cite{ref-roberts2}.

\begin{table}
\renewcommand*{\arraystretch}{1.15}

\begin{center}
\begin{tabular}{|c|c|}
\hline
Coefficent & Bound \\
\hline
$\Delta a^{X}=\frac{1}{2}\left(a_{d_{L}}^{X}+a_{d_{R}}^{X}-a_{s_{L}}^{X}-a_{s_{R}}^{X}
\right)$ & $10^{-21}$ GeV\\
$\Delta a^{Y}$ & $10^{-21}$ GeV\\
$\Delta a^{Z}$ & $10^{-17}$ GeV\\
$\Delta a^{T}$ & $10^{-16}$ GeV\\
\hline
\end{tabular}
\caption{
\label{table-oldbounds-a}
Strengths of the existing constraints on the CPT-violating differences between the $a$-type
coefficients for $d$ and $s$ quarks.
The values are taken from~\cite{ref-tables}, based on experimental kaon results
reported in~\cite{ref-nguyen,ref-babusci}.}
\end{center}
\end{table}

However, before we delve into questions about the $b$-type coefficients for quarks, we shall
consider a much less well studied area of the SME---the $a$-type coefficients for baryons. As pointed out in
section~\ref{sec-pion-CPTodd}, the $a^{\mu}$ are only observable as differences between
the coefficients for like-charged particles that can be interconverted. This immediately means that
to set any experimental bounds, it is necessary to go beyond two-flavor QCD; $a_{p}^{\mu}-a_{n}^{\mu}$
is not a QCD
observable, even in principle. We shall therefore extend our analysis to three-flavor QCD, with
a $s$ quark and assuming that there is a fairly robust $SU(3)_{f}$ symmetry. With this assumption,
the $a$-type coefficients for kaons as well as pions can be inferred from our formulas [as the kaons
are also pseudo-Goldstone bosons for the spontaneously broken $SU(3)_{L}\times SU(3)_{R}$; together
with the pions and the $\eta_{8}$, they form a flavor octet; we briefly discuss the extension of our
$\chi$PT methods to the $SU(3)_f$ sector in the appendix.]. Specifically, the kaon coefficient
is $a_{K^{0}}^{\mu}=a_{d}^{\mu}-a_{s}^{\mu}$ with no $s$-$d$ mixing.
Since the $K^{0}$ can oscillate into a
$\bar{K}^{0}$, it is possible to measure the difference of $a_{K^{0}}$ and $a_{\bar{K}^{0}}=-a_{K^{0}}$.
A number of strong bounds on the difference in quark coefficients, as measured in kaon oscillations
experiments, have been reported in the literature. The orders of magnitude of the best current constraints
are listed in table~\ref{table-oldbounds-a}.

What is remarkable is that, in the $SU(3)_{f}$ limit, the difference $a_{d}^{\mu}-a_{s}^{\mu}$ is the basis
of another observable:  the difference between the $a$-type coefficients for octet baryons that differ in their
valance quark content by the replacement of a $d$ quark with a $s$ quark. This means a difference such as
$a_{p}^{\mu}-a_{\Sigma^{+}}^{\mu}$, or the even more exotic $a_{\Sigma^{-}}^{\mu}-a_{\Xi^{-}}^{\mu}$.
The key relations follow from (\ref{eq-proton-a}) and its analogues for other species; these yield,
for example,
\begin{equation}
a_{p}^{\mu}-a_{\Sigma^{+}}^{\mu}=\frac{1}{2}
\left(a^{\mu}_{d_{L}}+a^{\mu}_{d_{R}}-a_{s_{L}}^{\mu}-a_{s_{R}}^{\mu}\right) .
\end{equation}
Conservative bounds [leaving at least an order of magnitude buffer to account for possible
deviations from $SU(3)_{f}$ symmetry] on such quantities are listed in table~\ref{table-newbounds-a}.

One thing that is notable about these bounds is that no method for constraining these baryon
coefficient differences has ever been proposed before! They would, in fact, be exceedingly difficult to
measure directly. (This is different from the situation with $a_{p}^{\mu}-a_{n}^{\mu}$ which is not
directly observable, even in principle---at least not without nonminimal couplings to gravity.)
Although baryons such as the proton and the $\Sigma^{+}$ can, in theory, interconvert
(there being no conserved quantity that differentiates them), the fact that there are
(in the standard model) no flavor-changing
neutral currents means that there can be no direct transitions between these species. What makes the
$K^{0}$-$\bar{K}^{0}$ system special is that the oscillation process is mediated by a box diagram that
exchanges both a $W^{+}$ and $W^{-}$, so that the net charges of the initial and final particles are the same.
There is no similar process for the baryons, so methods utilizing comparisons between different hadron
types represent essentially the only practicable way to constrain these differences.

\begin{table}
\renewcommand*{\arraystretch}{1.15}
\begin{center}
\begin{tabular}{|c|c|}
\hline
Coefficent & Bound \\
\hline
$a^{X}_{B}-a^{X}_{B'}$ & $10^{-20}$ GeV\\
$a^{Y}_{B}-a^{Y}_{B'}$ & $10^{-20}$ GeV\\
$a^{Z}_{B}-a^{Z}_{B'}$ & $10^{-16}$ GeV\\
$a^{T}_{B}-a^{T}_{B'}$ & $10^{-15}$ GeV\\
\hline
\end{tabular}
\caption{
\label{table-newbounds-a}
Order of magnitude bounds for differences between the $a$-type coefficients for $SU(3)_{f}$ octet baryons
$B$ and $B'$ that differ in quark content by one $d\leftrightarrow s$ replacement.}
\end{center}
\end{table}

The relations derived here from $\chi$PT can be used not just to place bounds on new combinations of
hadron SME parameters, but also on the underlying quark coefficients. This can be illustrated by
considering differences of nucleon $b$-type coefficients. According to
(\ref{eq-proton-b})---as well as the homologous formula for neutrons---
\begin{equation}
\label{eq-b-diff}
b^{\mu}_{p}-b^{\mu}_{n}=\frac{g_{A}}{2}\left(b^{\mu}_{u}-b^{\mu}_{d}\right),
\end{equation}
which contains no unknown LECs at LO in $\chi$PT.

There are bounds (coming from precision magnetometer experiments)
on linear combinations of mSME coefficients that include all
the proton and neutron spatial components $b^{J}_{p}$ and
$b^{J}_{n}$ ($J=X,Y,Z$), at $10^{-28}$--$10^{-33}$ GeV levels. With direct bounds on the proton and neutron
$b$-type terms, we could construct similarly precise bounds on the fundamental quark
parameters in (\ref{eq-b-diff}). Unfortunately however, the extant bounds are
actually on somewhat complicated linear combinations of
proton and neutron coefficients, including both
dimension-3 and dimension-4 terms. These mixtures of coefficients for operators of different
mass dimensions are unavoidable in purely nonrelativistic experiments, although it is possible to disentangle
the effects of, for
instance, $b^{J}$ and $d^{JT}$ at higher energies. In fact, this disentanglement can actually be accomplished
by using relativistic corrections related to nuclear binding and the internal motions of
constituent nucleons~\cite{ref-altschul39}, although separating the operators of different dimensions does
come with a significant cost in precision. The disentangled bounds will be worse than the raw
experimental ones by a sizable factor of $\sim m_{N}/\Delta e$, where $\Delta e$ is the difference in the
binding energies of the nucleons that are being probed in different nuclei.

However, to distinguish proton and neutron contributions, as
well as to separate di\-men\-sion-3 and dimension-4 operators, would require measurements of $b$-type
Lorentz violation for at least four different nuclear systems. At present, the best bounds on $b$-type
coefficients are dominated by measurements made on just two nuclei: $^{3}$He and
$^{129}$Xe~\cite{ref-brown2,ref-allmendinger},
which are very convenient to use
in atomic magnetometers, because they are spin-$\frac{1}{2}$ noble gasses. There is only one
other nucleus, $^{199}$Hg, for which comparably precise measurements have been made~\cite{ref-peck},
which means there are not enough independent measurements to extract complete and robust bounds on the
quark sector coefficients. However, naturalness does still suggest that the $b^{\mu}_{u}$ and $b^{\mu}_{d}$
should probably not be much larger than the best inferred bounds on $b^{\mu}_{p}$ and $b^{\mu}_{n}$.

\section{Conclusions and Outlook}
\label{sec-concl}

In this paper, we have given the first explorations of simultaneous quark and gluon SME operators of
dimension 4
in $\chi$PT, finding nontrivial relationships between the LECs that characterize their effects 
at the hadron level. We have also presented the first $\chi$PT analysis of dimension-3 SME operators.
The results for the dimension-3 CPT-violating terms have allowed us to place
new bounds on certain combinations of octet hadron $a$-type coefficients, based on comparisons to the
octet meson sector. This provides a novel avenue for constraining certain mSME parameters that are,
in principle, observable, but which would be extremely difficult to investigate directly.

In the course of our analyses, we have also made some additional observations about the character
of Lorentz-violating operators in $\chi$PT. 
There is a notable difference between the structure that $\chi$PT dictates for the CPT-even SME operators
(of dimension 4 and higher) and the CPT-odd ones (which begin at dimension 3). The dimension-4 terms
behave as modifications of the kinetic terms for the hadrons, and their sizes depend on the amount
of momentum carried by the individual quarks and gluons. There are nontrivial relations between the
coefficients for the PT-even quark-derived and gluon-derived terms.
The relations are tied to the physical 
fact that all the momentum of a given hadron must ultimately be carried by its constituent partons
(although those parton components generally
include sea quarks as well as valance quarks and gluons). However,
there are still a number of undetermined coefficients in the effective Lagrange densities for the
hadrons. These parameterize, for instance, the relative contributions from the isosinglet and isotriplet
Lorentz violation tensors, and they are ultimately determined by the interior wave functions
of the nucleons. Determination of the $\alpha^{(n)}$ and $\beta^{(n)}$ LECs, using nonperturbative
methods such as lattice QCD, would be a welcome development.

The situation is quite different for the dimension-3 operators, whose coefficients are, in the
chiral limit, completely determined by the transformation behavior of the quarks.
The Lorentz violation enters through
external fields that couple to the quarks, which means that the $l_{\mu}$, $r_{\mu}$, and $v_{\mu}^{(s)}$
terms contribute unambiguously to the pion and baryon effective actions. This also makes sense, since,
for example, the net $a$-type coefficient for a baryon will just be the sum of expectation values of the $a$-type
coefficients of its constituent quark fields. The contributions from the three valance quarks in a
$SU(3)_{f}$ octet baryon simply add up, while the contribution from the virtual sea of quark-antiquark
pairs cancels out.

There is, however, a subtlety to the $SU(3)_{f}$ analysis. For bounds that are based on kinematical
considerations---such as direction- and boost-dependent differences between the effective masses of
$K^{0}$ and $\bar{K}^{0}$ mesons---it is correct to phrase those bounds in terms of the mSME coefficients
(such as $a_{d}^{\mu}$ and $a_{s}^{\mu}$) for well-defined quark species. However, if the experimental
results are to be interpreted in terms of ``direct'' CPT violation---involving CPT-violating decays with
strangeness change $\Delta S=\pm1$, rather than asymmetric $K^{0}$-$\bar{K}^{0}$
oscillations involving $\Delta S=\pm2$---it would also be necessary to include in the analysis terms
such as $a_{ds}^{\mu}$, which parameterizes an operator
\begin{equation}
\label{eq-sd-mix}
{\cal L}^{d\,=\,3,\,{\rm CPT-odd}}_{s-d\,{\rm mixing}}=
-\frac{i}{2}\bar{s}a_{ds}^{\mu}\gamma_{\mu}d+{\rm h.\,c.},
\end{equation}
where ``h.\ c.'' indicates the hermitian conjugate. A term like (\ref{eq-sd-mix}), which is
off diagonal in flavor space, would contribute
directly to the kaon decay process, in an intrinsically Lorentz- and CPT-violating fashion. Whereas
the Cabibbo angle describes the mixing between the $s$ and $d$ species in the matrix of
the standard model's
fermion-Higgs Yukawa couplings, the $a_{ds}^{\mu}$ play analogous roles in the Lorentz-violating
sector. Further exploration of how neutral meson experiments could be used to place constraints on
$a_{ds}^{\mu}$ (as well as the other analogous mixing parameters that appear when more than three
flavors are taken into account) would be quite interesting.

In fact, it would also be useful to have systematic
methods for determining the effective SME coefficients for
heavier hadron species.
Using techniques for the study of hadrons containing heavy quarks ($c$ or $b$ flavors), it should
be possible to generalize the $\chi$PT results to answer questions about heavier mesons and the related
spin-$\frac{1}{2}$ baryons. The differences between the $a$-type coefficients for the constituents
of $D^{0}$ and $B^{0}$ mesons have already been measured, at roughly $10^{-15}$ GeV levels of precision.
These limits can presumably be translated into bounds on the differences of $a$-type coefficients
for baryons with the same heavy valance quarks.

It may also be possible to extend our analysis to mesons with spin. There has been some recent work
on higher-dimensional forms of Lorentz violation for spin-1 bosons~\cite{ref-mouchrek}. Lorentz
violation for a massive spin-1 particle is similar to that for a photon, although without the
restriction of gauge invariance there are additional allowed operators. The general features of
a Lorentz-violating mass term have been explored and appear to be qualitatively
understood~\cite{ref-gabadadze,ref-dvali,ref-altschul36}. If the mass-squared matrix $M^{\mu}\!{}_{\nu}$
for the vector boson field has an eigenvalue $m_{0}^{2}$ corresponding to a timelike direction and
a larger eigenvalue $m_{1}^{2}$ corresponding to a spacelike eigenvector, then there may
be propagation with signal and group velocities as large as $\frac{m_{1}}{m_{0}}>1$ for
the approximately longitudinal mode. However, in spite of these interesting results,
there has been no systematic survey of all possible Lorentz-violating operators of dimensions 3 and 4.

Existing work on Lorentz-invariant applications of $\chi$PT to spin-1 octet mesons, such
as in~\cite{ref-jenkins,ref-lutz1,ref-lutz2,ref-terschlusen1,ref-bruns,ref-terschlusen2},
has often focused on the forms taken by interaction vertices involving vector particles like the
$\rho^{0}$, rather than on the behavior of the vector propagator.
This focus is partially motivated by the vector meson dominance (VMD) phenomenon, in which
the interactions of hadrons with
deeply virtual photons can be dominated by diagrams in which the photon makes a virtual transition
into a neutral vector meson such as the $\rho^{0}$ before interacting with real hadrons.
Because of the existence of VMD, understanding the role of the vector meson sector of the SME may actually
be quite important for the interpretation of some high-energy collider tests of Lorentz and CPT symmetries.

Moreover, there are other heavy particles for which a different suite of techniques might be needed.
The $\chi$PT methodology has been useful for determining the effective Lorentz violation coefficients
for nucleons and pions. In terms of flavor $SU(3)_{f}$, these are the lightest representatives of the
meson and baryon octets. A natural additional question is how to determine the coefficients for
decuplet baryons as well. In fact, the mSME structure for a spin-$\frac{3}{2}$ field operator has not yet
been worked out, so even the general forms of the possible operators (much less their relationships
to the underlying quark and gluon operator structures) are unknown. The chief complication with a
spin-$\frac{3}{2}$ field is that the Rarita-Schwinger equation~\cite{ref-rarita} describes the behavior
of a field with both a Dirac index and a Lorentz index---and thus sixteen apparent components.
However, an actual spin-$\frac{3}{2}$ quantum has only eight possible states (four helicity projections,
along with a binary choice for particle versus antiparticle identity). Therefore only a certain
subspace of solutions
of the Rarita-Schwinger equation actually represents the propagation of spin-$\frac{3}{2}$ particles.
This significantly complicates the construction of any EFT theory for such particles; many of the operators
that might be constructed in generalizations of the Rarita-Schwinger Lagrange density will turn out
to be spurious (because they only affect the behavior of the unphysical part of the solution space) or
pathological (because they induce transitions between the physical subspace and the unphysical one, thus
destroying unitarity). This is a serious problem even for Lorentz-invariant
Rarita-Schwinger theories with nonminimal couplings~\cite{ref-velo,ref-porrati},
and it is likely to be an even greater challenge when the most general Lorentz-violating
couplings are included.
The inclusion of the $\Delta$ resonance in $\chi$PT in the Lorentz-invariant sector has been treated
extensively in the literature, addressing issues of power counting as well as the treatment of the
unphysical degrees of freedom, in such works as~\cite{Jenkins:1991es,Butler:1992pn,Hemmert:1997ye,
Pascalutsa:2002pi,Bernard:2003xf,Hacker:2005fh,Wies:2006rv}. Extensions of these methods to the
Lorentz-violating sector might be feasible.

In any event, understanding Lorentz violation for spin-$\frac{3}{2}$ composite particles such as $\Delta^{+}$
baryons would be very interesting, because of the importance of such particles to the
Greisen-Zatsepin-Kuzmin (GZK) cutoff~\cite{ref-greisen,ref-zatsepin}. Primary cosmic ray protons
of sufficient energy interact with cosmic microwave background photons according to
\begin{equation}
p^{+}+\gamma\rightarrow\Delta^{+}\rightarrow\left\{
\begin{array}{l}
p^{+}+\pi^{0} \\
n^{0}+\pi^{-}
\end{array}
\right.,
\end{equation}
and the threshold energy depends sensitively on the relevant $c$-type coefficient for the $\Delta^{+}$.
The process must be allowed for at least one $\Delta^{+}$ helicity state that is accessible from
each proton helicity state, in order for all the protons above the $\sim5\times10^{10}$ GeV GZK
threshold to have their energies drained away over intergalactic distances, as is observed
experimentally.

However, it is not even known how many different parameters actually govern the ultrarelativistic
dispersion relations for the $\Delta^{+}$ modes under the mSME. The propagation
of a field with spin-$\frac{3}{2}$ excitations may be controlled by up to four $c$-type
symmetric tensors, one for each helicity state. Alternatively, it may be that there are only two
independent tensors involved, with the $c$-type coefficients for a $\Delta^{+}$ taking the form
$c_{\Delta^{+}}^{\mu\nu}+2hd_{\Delta^{+}}^{\mu\nu}$, with $h$ being the helicity component of the
particle's angular momentum.

Either type of Lorentz-violating spin structure would be at least partially analogous
to the Lorentz-violating behavior of
relativistic spin-$\frac{1}{2}$ fermions, which have two helicity states and whose dispersion relations
are set by $c_{L}^{\mu\nu}=c^{\mu\nu}+d^{\mu\nu}$ and $c_{R}^{\mu\nu}=c^{\mu\nu}-d^{\mu\nu}$.
Note, however, that in spite of the Dirac spinor
having four components---allowing for the presence of two particle and two antiparticle excitation modes
for each momentum eigenvalue---there are not four separate $c$-type tensors, only the two. When the
C-parity of $\gamma_{5}$ is taken into account, the behavior of antiparticle modes is governed by the same
tensors as the particle modes. Something similar is expected for the spin-$\frac{3}{2}$ modes as well,
although the details of which Lorentz-violating terms actually change signs under the action of C
are unknown. (For relativistic fermion fields, regardless of their total spins, the {\em zitterbewegung}
process ensures that only helicity eigenstates are eigenstates of propagation. This ensures that the
even more complicated spin structure that is possible for Lorentz-violating integer-spin fields such as
photons---which is represented by the birefringent part of their bosonic Lagrange densities---cannot
be replicated for higher-spin fermions.)

Ultimately, although progress is being made in understanding the relationships between Lorentz violation
at the quark and gluon level and at the hadronic level, there are still important unanswered questions. As
$\chi$PT and other methods are used to further elucidate the connections between the SME coefficients
for different strongly-interacting particles, we expect there to be many strong new bounds based on
the understanding of these connections.

\section*{Acknowledgments}
This material is based upon work supported by the U.S. Department of Energy, Office of Science,
Office of Nuclear Physics, under Award Number DE-SC0019647 (MRS).

\appendix

\section*{Appendix: $SU(3)_f$ Formalism}

The extension of $\chi$PT methods to $SU(3)_f$ in the meson sector is straightforward. As in the $SU(2)$ case,
the Goldstone bosons are encoded in the matrix $U(x)$ of (\ref{eq-U}), which still transforms as in
(\ref{eq-U-transform}). However, the matrix $\phi$ in the exponential now takes the form 
\begin{equation}
\phi = \sum_{a=1}^8 \phi_a \lambda_a =\left[
\begin{array}{ccc}
\pi^0+\frac{1}{\sqrt{3}}\eta_{8} & \sqrt{2} \pi^+ & \sqrt{2} K^+ \\
\sqrt{2} \pi^- & -\pi^0+\frac{1}{\sqrt{3}}\eta_{8} & \sqrt{2} K^0 \\
\sqrt{2} K^- & \sqrt{2} \bar{K}^0 & -\frac{2}{\sqrt{3}}\eta_{8} 
\end{array}
\right],
\end{equation}
and the constant $F$ is now the pseudoscalar decay constant in the $SU(3)$ chiral limit---that is, with the
strange quark mass also set to zero.
Because the transformation properties are unchanged compared to the $SU(2)_f$ case, the LO Lagrange densities for
both the Lorentz-invariant and Lorentz-violating sectors still take the same forms as in (\ref{eq-LC-pion-LO}),
(\ref{eq-lo-pionlagrangian}), and (\ref{eq-d3-pion}), respectively. Differences between the two- and three-flavor
cases appear in the values of the low-energy constants, as well as possibly in the forms of higher-order
Lagrange densities, as some techniques used in reducing the number of independent terms at a given order (such as
the Caley-Hamilton formalism) may differ.

The extension to $SU(3)_f$ in the baryon sector is more complicated. Instead of the nucleon doublet $\Psi$,
the baryon octet is encoded in a traceless $3\times3$ matrix
\begin{equation}
B = \sum_{a=1}^8 B_a \frac{\lambda_a }{\sqrt{2}}=\left[
\begin{array}{ccc}
\frac{1}{\sqrt{2}}\Sigma^0 +  \frac{1}{\sqrt{6}} \Lambda & \Sigma^+ & p \\
\Sigma^- & -\frac{1}{\sqrt{2}}\Sigma^0 +  \frac{1}{\sqrt{6}} \Lambda & n \\
\Xi^- & \Xi ^ 0 & - \frac{2}{\sqrt{6}} \Lambda
\end{array}
\right],
\end{equation}
with the chiral transformation property
\begin{equation}
B \rightarrow K B K^\dagger.
\end{equation}
The corresponding covariant derivative is naively given by
\begin{equation}
D_\mu B = \partial_\mu B + [\Gamma_\mu,B].
\end{equation}
The Lagrangian is constructed by forming products of terms $X$ that each transform as $KXK^\dagger$
and then taking a trace. For example, the LO Lorentz-conserving meson-baryon Lagrange density is
\begin{equation}
\label{eq-octet-lag}
\mathscr{L}_{MB}^\text{LO} = \text{Tr}[\bar{B}(i\slashed{D}-m_0)B] -
\frac{D}{2}\text{Tr}\left(\bar{B}\gamma^\mu\gamma_5 \{ u_\mu, B \}\right) -
\frac{F}{2}\text{Tr}\left(\bar{B}\gamma^\mu\gamma_5 [ u_\mu, B ]\right).
\end{equation}
Here, $m_0$ is the octet baryon mass in the chiral limit, while $D$ and $F$ are LECs that can be
related to semi-leptonic decays. Note that there are three parameters, compared to two in the $SU(2)_f$ case.

Analogously, we expect the form of the Lorentz-violating Lagrange density in the $SU(3)_f$ sector to be
more complex. However, for the discussion in section~\ref{sec-expt}, we are only interested in the baryon
octet $a$-type coefficients. At LO, these enter through the covariant derivative term in (\ref{eq-octet-lag});
the terms proportional to $D$ and $F$ contribute to $b$-type terms, since they are proportional to $u_\mu$.
However, to properly include the Lorentz-violating interactions, the baryon covariant derivative has to be modified to
\begin{equation}
\label{eq-new-covdev}
D_\mu B = \partial_\mu B + [\Gamma_\mu,B] - i v_\mu^{(s)}B.
\end{equation}
In standard $\chi$PT, coupling to the vector current describes electromagnetic interactions, which at the
quark level are proportional to the quark charge matrix. Since this matrix is traceless, the singlet vector
current is identically zero. For the CPT-odd terms considered here, this is no longer the case, and the
$v_\mu^{(s)}$ contribution has to be considered. The $a$-type terms for the baryon octet can then be determined
from the first term in \eqref{eq-octet-lag}. In addition to reproducing the $SU(2)_f$ results of
section~\ref{sec-baryon-cpt-odd}, we find, for example, 
\begin{equation}
a^\mu_{\Sigma^+} = 2a_{u}^{\mu}+a_{s}^{\mu}.
\end{equation}

\end{document}